\documentclass[aps,prd,preprint,superscriptaddress,showpacs,longbibliography,footnoteadded]{revtex4-1}
\usepackage[T1]{fontenc}
\usepackage{float}
\usepackage{subfigure}
\usepackage{amssymb}
\usepackage{graphicx}
\usepackage{amsmath}
\usepackage{slashed}
\usepackage{tensor}
\usepackage{hyperref}
\usepackage{epstopdf}
\usepackage{extarrows}
\usepackage{color}
\usepackage{bbm}
\usepackage{caption}%caption 宏包，加强对caption的控制
\usepackage[section]{placeins}%可以禁止图片或者表格浮动超过\section的范围
\usepackage{makecell}%表格单元格内换行
\usepackage[utf8]{inputenc}
\hypersetup{
    colorlinks=true,
    linkcolor=red,
    citecolor=blue,
}

\begin{document}

\title{Massive Scalar Perturbations and Quasi-Resonance of  Rotating Black Hole in Analog Gravity}
\author{Hang Liu}
\email{hangliu@mail.nankai.edu.cn}
\affiliation{College of Physics and Materials Science, Tianjin Normal University, Tianjin 300387, China}

\author{Hong Guo}
\email{hong_guo@usp.br}
\affiliation{Escola de Engenharia de Lorena, Universidade de São Paulo, 12602-810, Lorena, SP, Brazil}

%%%%%%%%%%%%%%%%%%%%%%%%%%%%%%%%%%%%%%%%%%%%%%%%%%%%%%%%%%%%%%%%%%%%%%%%%%%%
\begin{abstract}
It was reported that the optical field fluctuations in self-defocusing media can be described by sound waves propagating in a two-dimensional photon-fluid which is controlled by the driving laser beam. The photon-fluid can be regarded  as the background where the sound waves propagate in the way like a  scalar field propagating in curved spacetime, thus providing a platform to study physics in analog gravity. In this analog gravity model, to be more specific, in the analog rotating black hole background, we study the quasinormal modes (QNMs) of massive scalar field perturbations, as a natural extension of the recent work on the massless QNMs in photon-fluid. We analyze the properties of the spectrum of fundamental QNMs which are calculated by Continued Fraction Method and WKB approximation method. We also investigate the  quasi-resonance  and find that it may exist in this analog gravity model. The existence of quasi-resonance is important to study the QNMs in analog gravity apparatus due to its slow damping rate and longevity  which give us better opportunity to probe and study it in laboratory.
\end{abstract}

%\date{\today}

\maketitle

%%%%%%%%%%%%%%%%%%%%%%%%%%%%%%%%%%%%%%%%%%%%%%%%%%%%%%%%%%%%%%%%%%%%%%%%%%%%
\section{Introduction}

Black holes, predicted as fundamental ``particles'' within the framework of general relativity, play the role as the crucial objects for understanding the evolution of the universe and probing the theories of gravity~\cite{Yagi:2016jml}. 
With the detection of gravitational waves produced by binary black hole mergers via ground-based gravitational wave detectors~\cite{LIGOScientific:2016lio,LIGOScientific:2016vlm,LIGOScientific:2018mvr}, gravitational wave astronomy has emerged as a significant tool for systematically investigating black hole properties and near-horizon physics~\cite{Barausse:2016eii,Barack:2018yly,Baibhav:2019rsa}. 
However, space-based gravitational wave detectors like LISA~\cite{LISA:2017pwj}, Taiji~\cite{Hu:2017mde,Gong:2021gvw} and TianQin~\cite{Gong:2021gvw,TianQin:2015yph}, which are capable of detecting low-frequency gravitational waves, will not be operational for several years. 
Meanwhile, the analysis of gravitational wave data still faces challenges such as immense computational demands and the need for a vast template database~\cite{Christensen:2022bxb,Thrane:2018qnx}. On the other hand, we also face huge challenges of  observing   some intriguing features of black holes in astrophysical environments, such as Hawking radiation, which is rather difficult to be observed, and such a difficulty prevents us from examining the validity of quantum field theory in curved spacetime. These obstacles delay the fulfillment of the growing demand for exploring black holes in the near term.

Due to the challenges in the detections of black holes and their relevant important characteristics arising from fundamental physics, it is fantastic if we can study black hole physics in laboratory by contemporary technology via the language of analogy, i.e. analog gravity. The theoretical side of this idea had been realized by Unruh in 1981, whose pioneering work \cite{Unruh:1980cg} laid the  basis of the analog gravity research over the past decades. Since then, people have made significant strides in this field. The analog studies of properties associated with astrophysical black holes have been extensively discussed, including mechanisms such as black hole superradiance~\cite{Basak:2002aw,Richartz:2009mi,Anacleto:2011tr,Patrick:2020baa}, and ongoing discussions on black hole thermodynamics~\cite{Zhang:2016pqx}, particularly topics like black hole entropy~\cite{Zhao:2012zz,Anacleto:2014apa,Anacleto:2015awa,Anacleto:2016qll}. 
In recent years, relativistic acoustic black holes have also been constructed in Minkowski spacetime using the Abel mechanism~\cite{Ge:2010wx,Anacleto:2010cr,Anacleto:2011bv,Anacleto:2013esa} and other methods~\cite{Bilic:1999sq,Fagnocchi:2010sn,Visser:2010xv,Giacomelli:2017eze}. 
Furthermore, researchers have explored acoustic black holes in curved spacetimes, including Schwarzschild spacetime~\cite{Guo:2020blq,Qiao:2021trw,Vieira:2021ozg,Ditta:2023lny} and Reissner-Nordström (RN) spacetime~\cite{Ling:2021vgk,Molla:2023hou}, as detailed in~\cite{Ge:2019our}.
On another front, the first acoustic black hole was successfully created in a rubidium Bose-Einstein condensate as early as 2009~\cite{Lahav:2009wx}. 
Following this, several key experiments~\cite{MunozdeNova:2018fxv,Isoard:2019buh} reported observations of thermal Hawking radiation in analog black holes and corresponding temperature measurements. 
For the latest research on analog Hawking radiation, see~\cite{Anacleto:2019rfn,Balbinot:2019mei,eskin2021hawking}. 
A series of foundational works~\cite{PhysRevA.70.063615,PhysRevA.69.033602,PhysRevLett.91.240407} laid the groundwork for studying analog gravity in ultracold quantum gases, which has seen further progress in recent years~\cite{Tian:2020bze,PhysRevA.106.053319,PhysRevD.105.124066,PhysRevD.107.L121502}. 
Additionally, an exciting recent breakthrough~\cite{Svancara:2023yrf} reported the observation of rotating curved spacetime characteristics within giant quantum vortices, further fueling interest in the study of analog gravity.

In modern black hole physics, quasinormal modes (QNMs) serve as a vital tool for probing information of black holes~\cite{Berti:2009kk,Konoplya:2011qq}. 
Linear perturbations around a perturbed black hole typically manifest as damped oscillations, where the real part of the QNM corresponds to the oscillation frequency, and the imaginary part determines the lifetime of the perturbation. 
During the ringdown phase of binary black hole mergers~\cite{Ma:2022wpv,Ghosh:2021mrv} and in extreme mass-ratio inspiral gravitational wave systems~\cite{Thornburg:2019ukt,Nasipak:2019hxh}, the QNM spectrum carries vital information about the black hole. 
Notably, QNMs are independent of the initial conditions of the perturbation, which makes them essential in gravitational wave observations and the study of astrophysical black holes~\cite{Franchini:2023eda}. Due to the importance of QNMs, it is natural and  significant to study QNMs in analog gravity.
Actually, QNMs have garnered significant attention in the study of analog gravity models. The theoretical studies on quasinormal modes (QNMs) can be traced back to early works on 2+1 and 3+1 dimensional acoustic black hole models~\cite{Visser:1997ux,Berti:2004ju,Cardoso:2004fi}, and recent advances in this area have also garnered widespread attention~\cite{Daghigh:2014mwa,Hegde:2018xub,Patrick:2020yyy,Liu:2024vde}. 
In 2018, Torres demonstrated that vortices in the scattering of water waves by rotating analog black holes could be fully characterized by QNMs~\cite{Torres:2017vaz}. 
This finding was further supported by experimental validation, where the QNMs of a 2+1 dimensional analog rotating black hole was studied~\cite{Torres:2020tzs}. 
Additionally, the excitation of QNMs in the Hawking radiation spectrum of analog black holes has been explored within quantum fluid analog models~\cite{Jacquet:2021scv}, highlighting the broader relevance of QNMs in both classical and quantum analog systems.
The study of QNMs in analog black holes allows for mutual validation through both theoretical and experimental approaches, thereby establishing a key research method for simulating and analyzing astrophysical black holes within the framework of analog gravity using laboratory platforms.

Recently, it was realized that rotating acoustic black holes can be generated within a self-defocusing optical cavity based on the fact that the fluid dynamics are also applicable to nonlinear optics~\cite{PhysRevA.78.063804}, thereby establishing a new platform which is called photon-fluid for analog black hole research.
The spin of the current analog black hole can be achieved by introducing a driving light beam with a vortex profile into the cavity.
Ever since the seminal work of~\cite{PhysRevA.78.063804}, this new analog gravity system has attracted much attentions in community, including research works on acoustic superradiance and superradiant instability~\cite{PhysRevA.80.065802,Ciszak:2021xlw}, experimental construction of this analog rotating black hole~\cite{Vocke2018}, measurement of superradiance in laboratory~\cite{Braidotti:2021nhw}, and the potential applications of the fluids system in the analog simulations of quantum gravity~\cite{Braunstein:2023jpo}. 
In addition, a recent paper~\cite{Krauss:2024vst} provided exciting insights about employing photon-fluid system to help resolve long-standing problems related to quantum gravity, including the black hole information-loss paradox and the removal of spacetime singularities.

These advancements have demonstrated that the photon-fluid system is a versatile and promising platform with significant potential to enhance our understanding of gravity through analogy.
In this paper, we focus on investigating the massive QNMs of an analog rotating black hole under the massive scalar field perturbations in a two-dimensional photon-fluid model.
As a natural extension of the recent work on the massless QNMs in photon-fluid~\cite{Liu:2024vde}, we examine the properties of fundamental QNM spectra, calculated using the Continued Fraction Method and WKB approximation. 
Additionally, we explore the possibility of quasi-resonance within this analog gravity model, noting its potential significance. 
The presence of quasi-resonance is particularly important for studying QNMs in analog gravity experiments, as its slow damping rate and extended duration provide a valuable opportunity for detailed investigation in a laboratory setting.
With the experimental advancements mentioned above, we are optimistic that the QNMs we have investigated here will soon be observable in apparatus of this photon-fluid model by which provides a novel testbed to black hole physics and theories of gravitation.

This work is organized as follows. 
In Sec.~\ref{sec2}, we derive the master equation of the fluctuation field. 
In Sec.~\ref{sec3}, we introduce the numerical methods employed in our calculations of QNF. 
In Sec.~\ref{sec4}, the numerical results of QNF are demonstrated and analyzed. 
In Sec.~\ref{sec5}, we investigate the existence of quasi-resonance.
The Sec.~\ref{sec6}, as the last section, is devoted to conclusions and discussions.

%%%%%%%%%%%%%%%%%%%%%%%%%%%%%%%%%%%%%%%%%%%%%%%%%%%%%%%%%%%%%%%%%%%%%%%%%%%%
\section{Basic Formulas}\label{sec2}

In this section, we briefly illustrate the basic parts of this photon-fluid system. 
The dynamics of the photon-fluid in our consideration is governed by the nonlinear Schrödinger equation (NSE) describing the slowly varying envelope of the optical field
\begin{equation}
\frac{\partial E}{\partial z}=\frac{i}{2k}\nabla^2 E-i\frac{k n_2}{n_0}E |E|^2,\label{ap_eq12}
\end{equation}
where $z$ is the coordinate of  propagation direction, $|E|^2$ stands for optical field intensity and $\nabla^2 E$ is defined by the transverse coordinates $(x, y)$, $k=2 \pi n_0/\lambda$ is the wave number, $\lambda$ is the laser wavelength in vacuum, $n_0$ is the linear refractive index, $n$ is the intensity dependent refractive index defined by $n=n_0-\Delta n=n_0-n_2|E|^2$ where  $n_2>0$ is the material  local nonlinear coefficient. Since the dynamics of this system lives on the transverse plane $(x,y)$, the propagation coordinate $z$ acts as  effective time by relation $t=\frac{n_0}{c}z$. In addition to the local nonlinearity, a non-local thermo-optical nonlinearities can be introduced to some medium such that \cite{Marino:2019flp,Ciszak:2021xlw}
\begin{equation}
\Delta n=n_2 |E|^2+n_{th}.\label{nonlocal}
\end{equation}

Based on NSE, the following effective metric of an analog rotating black hole can be obtained \cite{PhysRevA.78.063804,PhysRevA.80.065802,Ciszak:2021xlw}
\begin{equation}
d s^{2} =-\left(1-\frac{r_{H}}{r}-\frac{r_{H}^{4} \Omega_{H}^{2}}{r^{2}}\right) d t^{2}+\left(1-\frac{r_{H}}{r}\right)^{-1} d r^{2}-2 r_{H}^{2} \Omega_{H} d \theta d t+r^{2} d \theta^{2},
\end{equation}
where $r_H$ is the event horizon, $\Omega_H$ represents the angular velocity of the black hole, $\xi$ describes the healing length, $j$ is an integer representing the topological charge of the optical vortices, and they are given by
\begin{equation}
r_{H}=\frac{\xi^{2}}{r_{0}}, \quad \Omega_{H}=\frac{j \xi}{\pi r_{H}^{2}}, \quad \xi=\frac{\lambda}{2 \sqrt{n_{0} n_{2} \rho_{0}}}.
\end{equation}
In this analog black hole back ground,  the propagation  of  density perturbation $\rho_1$ of optical field is governed by a massive Klein-Gordon equation \cite{Ciszak:2021xlw}
\begin{equation}
	\Box \rho_1-\mu^2\rho_1=\frac{1}{\sqrt{-g}}\partial_{\mu}(\sqrt{-g}g^{\mu\nu}\partial_{\mu}\rho_1)-\mu^2\rho_1=0,
\end{equation}
where the mass term $\mu^2$ comes from the introduction of the non-local thermo-optical nonlinearities $n_{th}$ introduced by Eq.~\eqref{nonlocal}, and $g$ is the determinant of the effective metric. With the massive Klein-Gordon equation, we can obtain the master equation of the perturbation field
\begin{align}
&\frac{d^2\Psi(r_\ast)}{dr^2_\ast}+U(\omega,r)\Psi(r_\ast)=0\label{mastereq},\\
&U(\omega,r)=\left(\omega-\frac{m\Omega_H}{r^2}\right)^2-\left(1-\frac{1}{r}\right)\left(\frac{m^2}{r^2}+\frac{1}{2r^3}-\frac{1}{4r^2}\left(1-\frac{1}{r}\right)+\mu^2\right),
\end{align}
where we have set $r_H=1$ which means that $r$ is measured in units of $r_H$, and both $\omega$ and $\Omega_H$ are measured in units of $r_H^{-1}$. The derivation of this equation can be found in Appendix \ref{app1}.
Then for our  asymptotically flat analog black hole spacetime, the QNF, i.e. $\omega$,  can be determined by solving  Eq.~\eqref{mastereq} associated with the following boundary conditions
\begin{equation}
\Psi \sim
\begin{cases}
   e^{-i(\omega-m \Omega_H) r_\ast}, &  r_\ast \to -\infty\quad(r\to r_H,r_H=1), \\
   e^{+i\sqrt{\omega^2-\mu^2} r_\ast}, &  r_\ast \to +\infty\quad(r\to +\infty),
\end{cases}
\label{master_bc0}
\end{equation}
which means that the perturbation field is ingoing at the event horizon and outgoing at infinity.
%%%%%%%%%%%%%%%%%%%%%%%%%%%%%%%%%%%%%%%%%%%%%%%%%%%%%%%%%%%%%%%%%%%%%%%%%%%%
\section{The Methods}\label{sec3}

In this section, we would like to introduce two different numerical methods for calculating QNF of this analog black hole. As a standard procedure, the QNF is usually  calculated by one method and then confirmed by another one, since this process can not only  enhance the reliability of our numerical results but also examine the validity of the numerical methods.

%%%%%%%%%%%%%%%%%%%%%%%%%%%%%%%%%%%%%%%%%%%%%%%%%%%%%%%%%%%%%%%%%%%%%%%%%%%%
\subsection{Leaver's Continued Fraction Method (CFM)}

Leaver~\cite{Leaver,PhysRevD.41.2986} first  calculated QNF  by numerically solving a three-term recurrence relation, which is well-known as CFM. To learn more about this method please refer to \cite{Leaver,PhysRevD.41.2986,Konoplya:2011qq} for a comprehensive introduction.

In current context, it is better to work in the original radial coordinate, under which the master equation takes the form
\begin{equation}
\Psi''(r)+\frac{1}{r(r-1)}\Psi'(r)+\frac{r^2}{(r-1)^2}U(\omega,r)\Psi(r)=0,\label{eq4}
\end{equation}
which has two regular singular points at $r=1$ and $r=0$, and one irregular singularity at $r\to\infty$. According to the asymptotic behavior (boundary condition Eq.~\eqref{master_bc0}) of the perturbation field $\Psi$, we are able to get the asymptotic solutions at the horizon $r\to r_H$ and infinity $r\to\infty$ in the original radial coordinate. First, when $r\to r_H$, we have ingoing modes
\begin{equation}
\Psi(r)\thicksim (r-1)^{-i(\omega-m\Omega_H)},\quad r\to r_H.	\label{bc_rh}
\end{equation}
While when $r\to\infty$, we need to be carful to get the appropriate  asymptotic solution due to the fact that infinity is a irregular singularity, which indicates that beside the dominant exponential behavior, the subdominant power law behavior should also be considered in order to make the CFM as accurate as possible. To this end, we take the following ansatz of outgoing modes at infinity
\begin{equation}
\Psi(r)\thicksim	 e^{i\sqrt{\omega^2-\mu^2}r}r^{\kappa}.
\end{equation}
Substituting this formula back to Eq.~\eqref{eq4} and take a limit $r\to\infty$, we can get 
\begin{equation}
\kappa=\frac{i(2\omega^2-\mu^2)}{2\sqrt{\omega^2-\mu^2}},	
\end{equation}
which leads us to 
\begin{equation}
\Psi(r)\thicksim	 e^{i\sqrt{\omega^2-\mu^2}r}r^{\frac{i(2\omega^2-\mu^2)}{2\sqrt{\omega^2-\mu^2}}},\quad r\to\infty. \label{bc_rinf}
\end{equation}
Combining the asymptotic solutions  which are divergent at the boundary, and then expanding the remaining part  into the Frobenius series around event horizon $r=1$, we arrive at an appropriate expansion formulation of $\Psi(r)$
\begin{equation}
	\Psi(r)=e^{i\sqrt{\omega^2-\mu^2} r}r^{i\left(\sqrt{\omega^2-\mu^2}+\frac{\mu^2}{2\sqrt{\omega^2-\mu^2}}\right)+i(\omega-m\Omega_H)}(r-1)^{-i(\omega-m\Omega_H)}\sum_{n=0}^{\infty}a_n \left(\frac{r-1}{r}\right)^n.\label{expansion}
\end{equation}
The merit of this expansion is that the singular factor (i.e. boundary conditions for QNMs) of the solution to master equation has been  singled out in front of a series which is required to be convergent in the region $1\leq r\leq\infty$, and this convergence can only be achieved by QNF. In addition, one can check that the boundary conditions Eq.~\eqref{bc_rh} and Eq.~\eqref{bc_rinf} are indeed satisfied by Eq.~\eqref{expansion}.

The next step is to derive the recurrence relation of the expansion coefficients $a_n$. 
To this end, we just need to substitute Eq.~\eqref{expansion} into Eq.~\eqref{eq4}, but before we do this, it seems necessary to work in a new coordinate 
\begin{equation}
z=\frac{r-1}{r},
\end{equation}
by which we have
\begin{equation}
	\Psi(r)=e^{i\sqrt{\omega^2-\mu^2} r}r^{i\left(\sqrt{\omega^2-\mu^2}+\frac{\mu^2}{2\sqrt{\omega^2-\mu^2}}\right)}z^{-i(\omega-m\Omega_H)}\sum_{n=0}^{\infty}a_n z^n.\label{expansion2}
\end{equation}
Within this coordinate, we can get the following three-term recurrence relation
\begin{equation}
\begin{split}
&\alpha_0 a_1+\beta_0 a_0=0,\\
&\alpha_n a_{n+1}+\beta_n a_n+\gamma_n a_{n-1}=0, \quad n\geq1,
\end{split}
\end{equation}
where
\begin{equation}
\begin{aligned}
\alpha_n &=4(1+n)(\mu-\omega)(\mu+\omega)\left(1+n-2 i \omega+2 i m \Omega_H\right), \\
 \beta_n &=-4 \mu^4+2 \mu^2\left(-2 m^2-(1+2 n)^2+4 i(1+2 n) \omega+10 \omega^2+3 i \sqrt{-\mu^2+\omega^2}+6 i n \sqrt{-\mu^2+\omega^2}\right. \\
&\left.+6 \omega \sqrt{-\mu^2+\omega^2}\right)-2 \omega^2\left(-1-2 m^2-4 n^2+4 i \omega+4 i \sqrt{-\mu^2+\omega^2}+8 \omega\left(\omega+\sqrt{-\mu^2+\omega^2}\right)\right. \\
&\left.+4 i n\left(i+2 \omega+2 \sqrt{-\mu^2+\omega^2}\right)\right)+4 m\left(-3 \mu^2 \sqrt{-\mu^2+\omega^2}+4 \omega^2 \sqrt{-\mu^2+\omega^2}\right. \\
&\left.\quad+2(i+2 i n+2 \omega)\left(-\mu^2+\omega^2\right)\right) \Omega_H, \\
\gamma_n & =\mu^4+\omega^2\left(1-4 n^2+8 i n\left(\omega+\sqrt{-\mu^2+\omega^2}\right)+8 \omega\left(\omega+\sqrt{-\mu^2+\omega^2}\right)\right) \\
& -\mu^2\left(1-4 n^2+4 i n\left(2 \omega+\sqrt{-\mu^2+\omega^2}\right)+4 \omega\left(2 \omega+\sqrt{-\mu^2+\omega^2}\right)\right) \\
& +4 m\left(2(\mu-\omega)(i n+\omega)(\mu+\omega)+\mu^2 \sqrt{-\mu^2+\omega^2}-2 \omega^2 \sqrt{-\mu^2+\omega^2}\right) \Omega_H.
\end{aligned}
\end{equation}
With this three-term recurrence relation, the ratio of successive $a_n$ can be formulated  in two ways, one is given by infinite continued fraction as
\begin{equation}
\frac{a_{n+1}}{a_n}=\frac{-\gamma_{n+1}}{\beta_{n+1}-\frac{\alpha_{n+1} \gamma_{n+2}}{\beta_{n+2}-\frac{\alpha_{n+2} \gamma_{n+3}}{\beta_{n+3}-...}}},
\end{equation}  
another one is by finite continued fraction
\begin{equation}
\frac{a_{n+1}}{a_n}=\frac{\gamma_n}{\alpha_n} \frac{\alpha_{n-1}}{\beta_{n-1}-\frac{\alpha_{n-2} \gamma_{n-1}}{\beta_{n-2}-\alpha_{n-3} \gamma_{n-2} / \ldots}}-\frac{\beta_n}{\alpha_n}.
\end{equation}
Thus the equivalence relation between the two equations above  can be employed to allow us to numerically obtain infinite number of roots corresponding to QNF of the following equation for $n\geq1$

\begin{equation}
\frac{\gamma_{n+1}}{\beta_{n+1}-\frac{\alpha_{n+1} \gamma_{n+2}}{\beta_{n+2}-\frac{\alpha_{n+2} \gamma_{n+3}}{\beta_{n+3}-...}}}=\frac{\beta_n}{\alpha_n}-\frac{\gamma_n}{\alpha_n} \frac{\alpha_{n-1}}{\beta_{n-1}-\frac{\alpha_{n-2} \gamma_{n-1}}{\beta_{n-2}-\alpha_{n-3} \gamma_{n-2} / \ldots}}, \quad n\geq 1\label{icf}
\end{equation}  
and for $n=0$
\begin{equation}
\beta_0-\frac{\alpha_0 \gamma_1}{\beta_1-\frac{\alpha_1 \gamma_2}{\beta_2^{\prime}-\frac{\alpha_2 \gamma_3}{\beta_3-\ldots}}}=0. \label{cf}
\end{equation}
Actually, for every $n\geq0$, Eq.~\eqref{icf} and Eq.~\eqref{cf} are completely equivalent in the sense that every solution to Eq.~\eqref{icf} is also a solution to Eq.~\eqref{cf}, and vice versa. However, the most stable roots depend on $n$, i.e. the $n$th QNF $\omega_n$ is  numerically the most stable roots of Eq.~\eqref{icf} for a certain $n>0$ (overtone modes)  or Eq.~\eqref{cf}  for $n=0$ (fundamental modes) \cite{Leaver,PhysRevD.41.2986}.

%%%%%%%%%%%%%%%%%%%%%%%%%%%%%%%%%%%%%%%%%%%%%%%%%%%%%%%%%%%%%%%%%%%%%%%%%%%%
\subsection{WKB Approximation Method}

WKB approximation method is also a powerful method in the calculation of QNF. 
By WKB, the QNF can be determined by solving following equation~\cite{Konoplya:2019hlu}
\begin{equation}
\begin{aligned}
\omega^2&=V_0(\omega)+A_2\left(\mathcal{K}^2,\omega\right)+A_4\left(\mathcal{K}^2,\omega\right)+A_6\left(\mathcal{K}^2,\omega\right)+\ldots\\
&-i \mathcal{K} \sqrt{-2 V_2(\omega)}\left(1+A_3\left(\mathcal{K}^2,\omega\right)+A_5\left(\mathcal{K}^2,\omega\right)+A_7\left(\mathcal{K}^2,\omega\right) \ldots\right),
\end{aligned}\label{eq10}
\end{equation}
where potential $V(\omega,r)$ satisfies $U(\omega,r)=\omega^2-V(\omega,r)$, and $V_m(\omega)$ are given by
\begin{equation}
V_{m}=\left.\frac{d^mV(\omega,r_\ast)}{dr_\ast^m}\right|_{r_\ast=r_{\ast o}},\quad m\geq2,
\end{equation}
in which $r_{\ast o}$ is the location of the maximum value of $V(\omega,r_\ast)$ in tortoise coordinate, and naturally $V_0=V(\omega,r_{\ast o})$.
$A_{k}(\mathcal{K}^2,\omega)$ are polynomials of $V_2,V_3,\ldots V_{2k}$, and $\mathcal{K}$ stands for eikonal formula
\begin{equation}
\mathcal{K}=i\frac{\omega^2-V_0}{\sqrt{-2V_2}}.
\end{equation}
By employing  the boundary conditions of QNMs, it is found that $\mathcal{K}$ has to satisfy 
\begin{equation}
\mathcal{K}=n+\frac{1}{2},\quad n\in\mathbb{N},	
\end{equation}
where $n$ is the overtone number. 
With these basic  formulas of WKB, we are going to calculate QNF with $6$th order WKB approximation method, which means that we have to calculate $A_k$ up to $A_6$. 
The expressions of $A_k$ are rather complex and we just only demonstrate $A_2$ and $A_3$ as examples 
\begin{equation}
\begin{aligned}
A_3&=\frac{1}{13824 V_2^5}\Bigg[-940\left(n+\frac{1}{2}\right)^2 V_3^4+1800\left(n+\frac{1}{2}\right)^2 V_2 V_4 V_3^2-672\left(n+\frac{1}{2}\right)^2 V_2^2 V_5 V_3\\
& -204\left(n+\frac{1}{2}\right)^2 V_2^2 V_4^2+96\left(n+\frac{1}{2}\right)^2 V_2^3 V_6-385 V_3^4+918 V_2 V_4 V_3^2-456 V_2^2 V_5 V_3 \\
&  -201 V_2^2 V_4^2+120 V_2^3 V_6\Bigg],
\end{aligned}
\end{equation}
and 
\begin{equation}
	A_2=\frac{-60\left(n+\frac{1}{2}\right)^2 V_3^2+36\left(n+\frac{1}{2}\right)^2 V_2 V_4-7 V_3^2+9 V_2 V_4}{288 V_2^2}.
\end{equation}
%%

%%%%%%%%%%%%%%%%%%%%%%%%%%%%%%%%%%%%%%%%%%%%%%%%%%%%%%%%%%%%%%%%%%%%%%%%%%%%
\section{Numerical Results of Massive QNF}\label{sec4}

In this section we demonstrate and analyze the properties of the numerical results of QNF of massive scalar perturbation. To obtain the QNF, we adopt previously introduced CFM and WKB method. In the cases of winding number $m>0$, both CFM and WKB are used and they give consistent results, while in the $m<0$ case we only use CFM due to the poor performance of WKB for massive QNF.

In Table~\ref{tab1} we show the numerical results of the fundamental QNF at winding number $m=1$ for different values of mass $\mu$ and angular velocity $\Omega_H$. The data displayed in this table are obtained by CFM and WKB  whose results are in good agreement with each other, which confirms  the validity  of the numerical methods we adopted. For the QNF's real part $\omega_R$ which stands for the oscillation frequency of QNMs, it can be observed that $\omega_R$ will monotonically increase  with the grow of $\Omega_H$, and this behavior holds for all the mass values considered in this table. On the hand,  one can find that for any given $\Omega_H$, the $\omega_R$ can be also improved by increasing the value of mass $\mu$. These results suggest that QNMs with higher oscillation frequency can be observed either by choosing a background driving  optical field whose profile can induce faster rotating analog black hole, or choosing the medium of which  non-local thermo-optical nonlinearities can bring us bigger effective  mass of perturbation field. For the QNF's imaginary part $\omega_I$ which is negative and reflects the damping rate of QNMs, we find that its magnitude becomes larger when increasing $\Omega_H$, and  this property holds for all scalar perturbations with different mass in our current consideration. This result indicates that QNMs will have a shorter life in this analog rotating black hole spacetime with higher angular velocity $\Omega_H$, since  a negative $\omega_I$ with larger magnitude clearly suggests a faster damping rate of QNMs. In contrast, it is found that the magnitude of $\omega_I$ can be decreased  by increasing the mass value $\mu$ of the perturbation field, such that for a given $\Omega_H$ the QNMs can live longer if it has a bigger effective mass. Furthermore, the behavior of $\omega_R$ and $\omega_I$ under the change of $\Omega_H$ manifests a remarkable  contrast,  as $\omega_R$ monotonically grows with the increment of $\Omega_H$ and seems to be unbounded, while for $\omega_I$ whose magnitude can be also monotonically improved by higher $\Omega_H$ but all $\omega_I$ seems to approach a constant around $\omega_I\approx-0.25$ when $\Omega_H$ is large enough. Additionally, the differences between QNF, both real and imaginary part, induced by different mass values, will be suppressed by sufficient large  $\Omega_H$, as a natural consequence of  the effects of $\mu$ on QNF can be ignored when compared with the effects of large $\Omega_H$.

%%%%%%%%%%%%%%%%%%%%%%%%%%%%%%%
\begin{table}[!htbp]
\centering
\caption*{$m=1$} 
\resizebox{\textwidth}{!}
{
        \begin{tabular}{ccccccc}
    \hline\hline
    $\mu$ &Method&  $\Omega_H=0$  &$\Omega_H=0.5$          & $\Omega_H=1$          & $\Omega_H=5$             & $\Omega_H=10$       \\
    \hline
    $0$ &WKB&     $ 0.366152 -0.194512i $           & $ 0.676862 - 0.222138i $ & $ 1.10283 - 0.237139i $ & $ 5.02137 - 0.249731i $    & $ 10.0119 - 0.249279i $\\
        \cline{2-7}
        &CFM&     $ 0.365926 -0.193965i $           & $ 0.676790 -0.222215i $ & $ 1.10290 -0.237129i $ & $ 5.02182 -0.249306i $    & $ 10.0109 -0.249825i $\\
        \hline
    $0.1$ &WKB&   $ 0.368812 - 0.191628i $           & $ 0.678529 - 0.221165i $ & $ 1.10403 - 0.236772i $ & $ 5.02260 - 0.248940i $    & $ 10.0112 - 0.249950i $\\
        \cline{2-7}
        &CFM&     $ 0.368582 -0.191090i $           & $ 0.678460 -0.221261i $ & $ 1.10398 -0.236755i $ & $ 5.02207 -0.249288i $    & $ 10.0111 -0.249820i $\\
        \hline
    $0.3$ &WKB&   $0.389847 - 0.167652i $           & $ 0.691903 - 0.213403i $ & $ 1.11275 - 0.233745i $ & $ 5.02405 - 0.249065i $    & $ 10.0115 - 0.249943i $\\
        \cline{2-7}
        &CFM&     $ 0.389569 -0.167218i $           & $ 0.691830 -0.213516i $ & $ 1.11266 -0.233748i $ & $ 5.02405 -0.249139i $    & $ 10.0121 -0.249783i $\\
        \hline
    $0.5$ &WKB&   $ 0.416878 - 0.111628i $           & $ 0.718672 - 0.197276i $ & $ 1.13004 - 0.227639i $ & $ 5.02811 - 0.248758i $    & $ 10.0138 - 0.249807i $\\
        \cline{2-7}
        &CFM&     $ 0.428464 -0.115165i $           & $ 0.718603 -0.197413i $ & $ 1.13008 -0.227633i $ & $ 5.02803 -0.248842i $    & $ 10.0141 -0.249708i $\\
        \hline\hline
\end{tabular}
}
\caption{The fundamental QNF obtained by WKB and CFM  at $m=1$ for different mass values $\mu$ and angular velocity $\Omega_H$.}\label{tab1}
\end{table}
%%%%%%%%%%%%%%%%%%%%%%%%%%%%%%%

%%%%%%%%%%%%%%%%%%%%%%%%%%%%%%%
\begin{figure}[thbp]
\centering
\includegraphics[height=2.4in,width=3.2in]{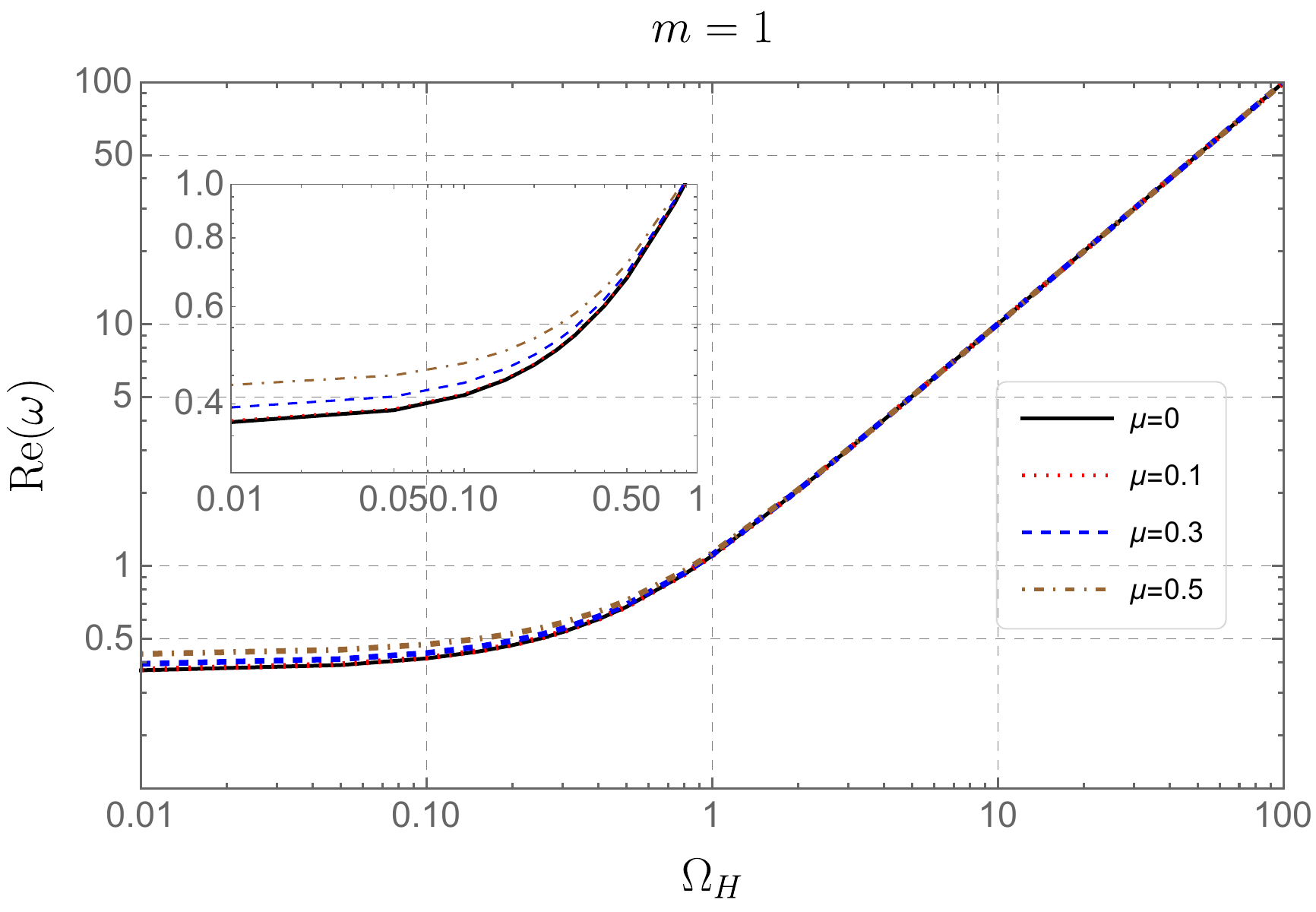}
\includegraphics[height=2.4in,width=3.2in]{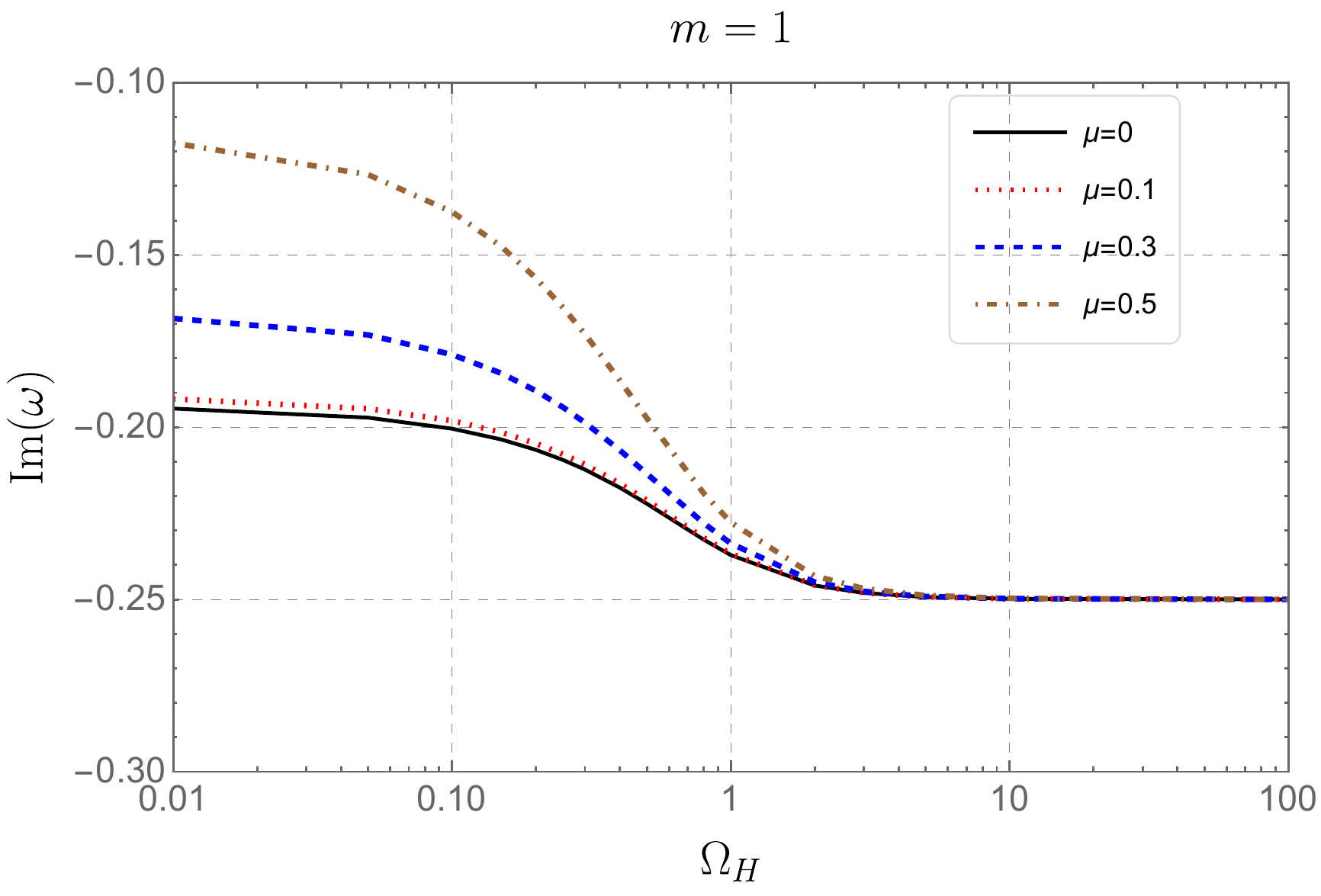}
\caption{The dependence of real part $\omega_R$ (left plot) and imaginary part $\omega_I$ (right plot) of fundamental QNF on $\Omega_H$ at $m=1$ for various mass values.\label{fig1}}
\end{figure}
%%%%%%%%%%%%%%%%%%%%%%%%%%%%%%%

%%%%%%%%%%%%%%%%%%%%%%%%%%%%%%%
\begin{table}[!htbp]
\centering
\caption*{$m=5$} 
\resizebox{\textwidth}{!}
{
        \begin{tabular}{ccccccc}
    \hline\hline
    $\mu$ &Method&  $\Omega_H=0$  &$\Omega_H=0.5$          & $\Omega_H=1$          & $\Omega_H=5$             & $\Omega_H=10$       \\
    \hline
    $0$ &WKB&     $ 1.92059 - 0.192508i $           & $ 3.45759 - 0.223514i $ & $ 5.57214 - 0.238331i $ & $ 25.1239 - 0.24938i $    & $ 50.0621 - 0.249844i $\\
        \cline{2-7}
        &CFM&     $ 1.92059 -0.192507i $           & $ 3.45759 -0.223514i $ & $ 5.57214 -0.238330i $ & $ 25.1239 -0.24938i $    & $ 50.0621 -0.249844i $\\
        \hline
    $0.5$ &WKB&     $ 1.94189 - 0.188920i $           & $ 3.46769 - 0.222449i $ & $ 5.578 - 0.237937i $ & $ 25.1246 - 0.247791i $    & $ 50.0606 - 0.253713i $\\
        \cline{2-7}
        &CFM&     $ 1.94189 -0.188919i $           & $ 3.46769 -0.222449i $ & $ 5.578 -0.237936i $ & $ 25.1252 -0.249362i $    & $ 50.0628 -0.249839i $\\
        \hline
    $1.5$ &WKB&     $ 2.11545 - 0.159159i $           & $ 3.54898 - 0.213791i $ & $ 5.62506 - 0.234771i $ & $ 25.1352 - 0.249266i $    & $ 50.0638 - 0.258024i $\\
        \cline{2-7}
        &CFM&     $ 2.11545 -0.159159i $           & $ 3.54898 -0.213791i $ & $ 5.62506 -0.234770i $ & $ 25.1351 -0.249213i $    & $ 50.0678 -0.249801i $\\
        \hline
    $2.5$ &WKB&     $ 2.48379 - 0.0896973i $           & $ 3.71443 - 0.195614i $ & $ 5.71977 - 0.228318i $ & $ 25.1552 - 0.249032i $    & $ 50.0779 - 0.249314i $\\
        \cline{2-7}
        &CFM&     $ 2.48381 -0.0897163i $           & $ 3.71443 -0.195614i $ & $ 5.71977 -0.228318i $ & $ 25.1551 -0.248915i $    & $ 50.0777 -0.249727i $\\
        \hline\hline
\end{tabular}
}
\caption{The fundamental QNF obtained by WKB and CFM  at $m=5$ for different mass values $\mu$ and angular velocity $\Omega_H$.}\label{tab2}
\end{table}
%%%%%%%%%%%%%%%%%%%%%%%%%%%%%%%

\begin{figure}[thbp]
\centering
\includegraphics[height=2.4in,width=3.2in]{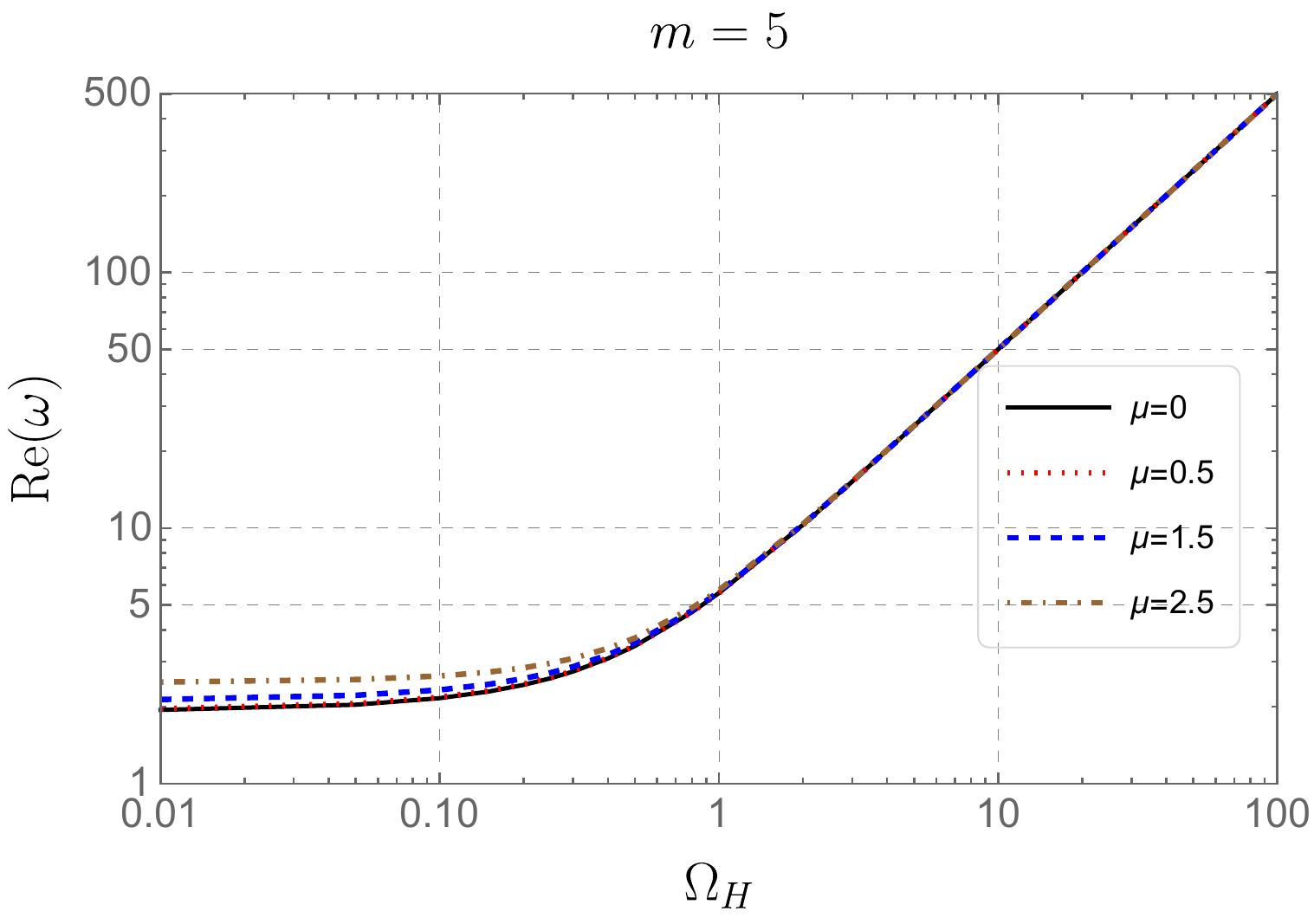}
\includegraphics[height=2.4in,width=3.2in]{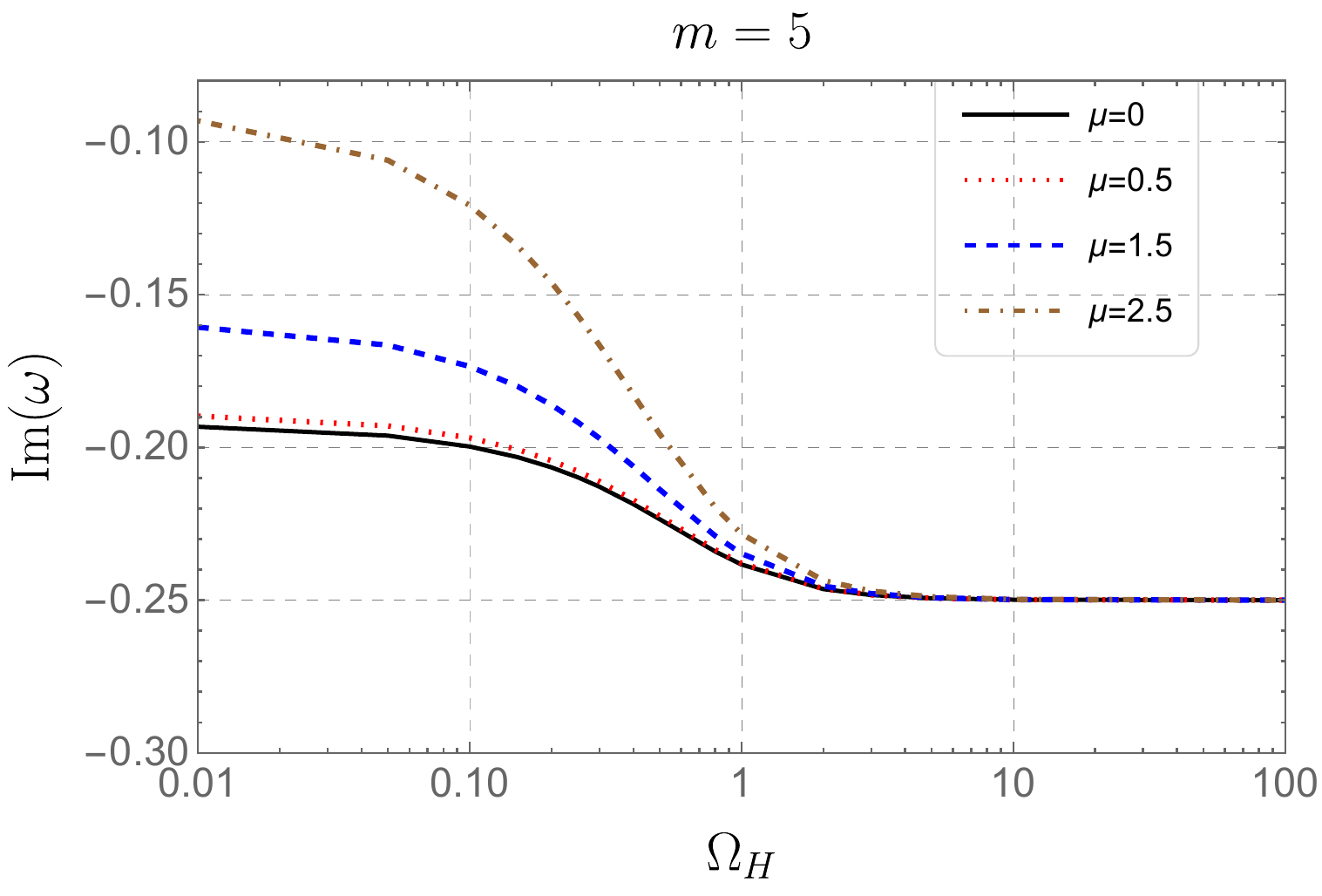}

\includegraphics[height=2.4in,width=3.2in]{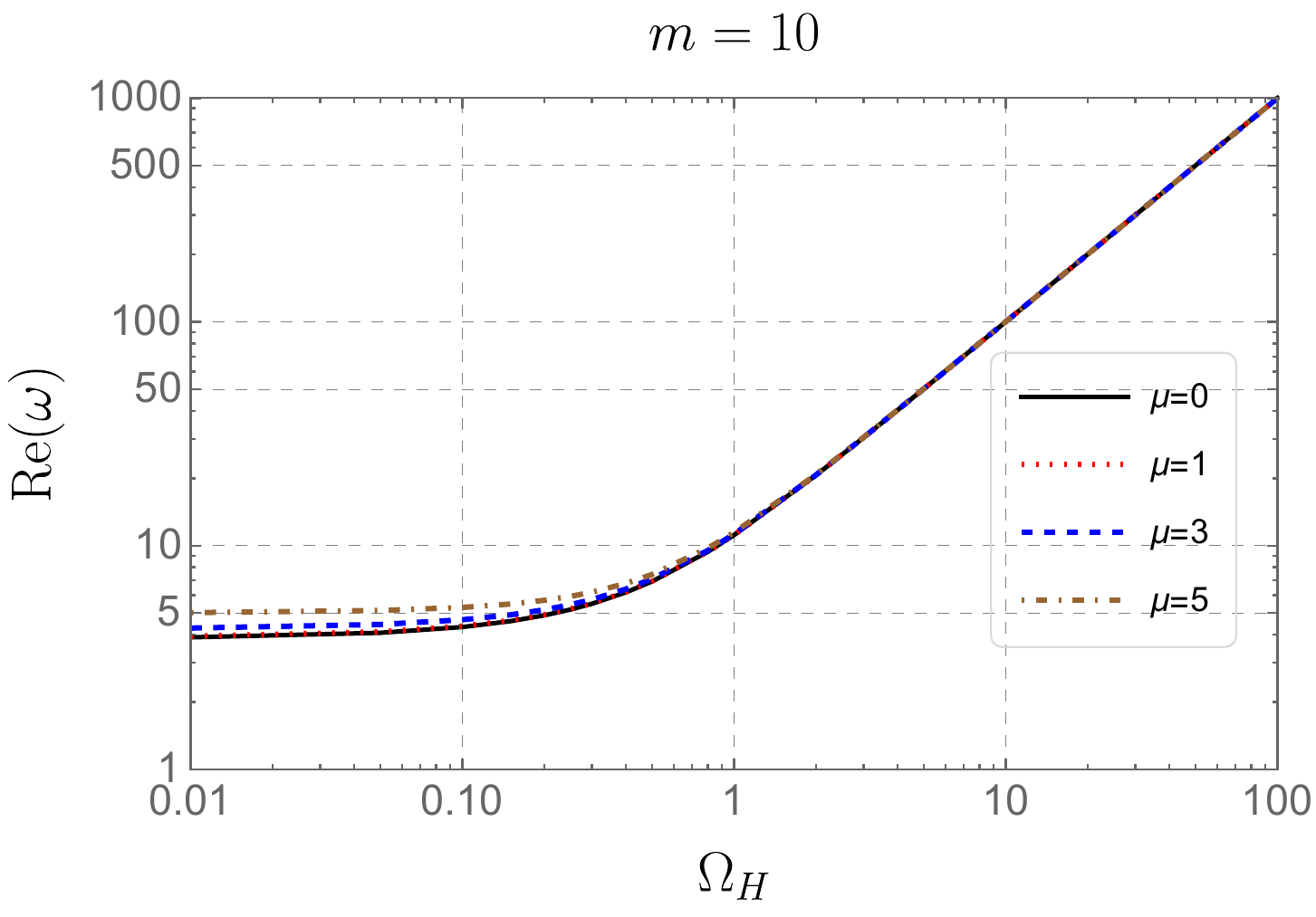}
\includegraphics[height=2.4in,width=3.2in]{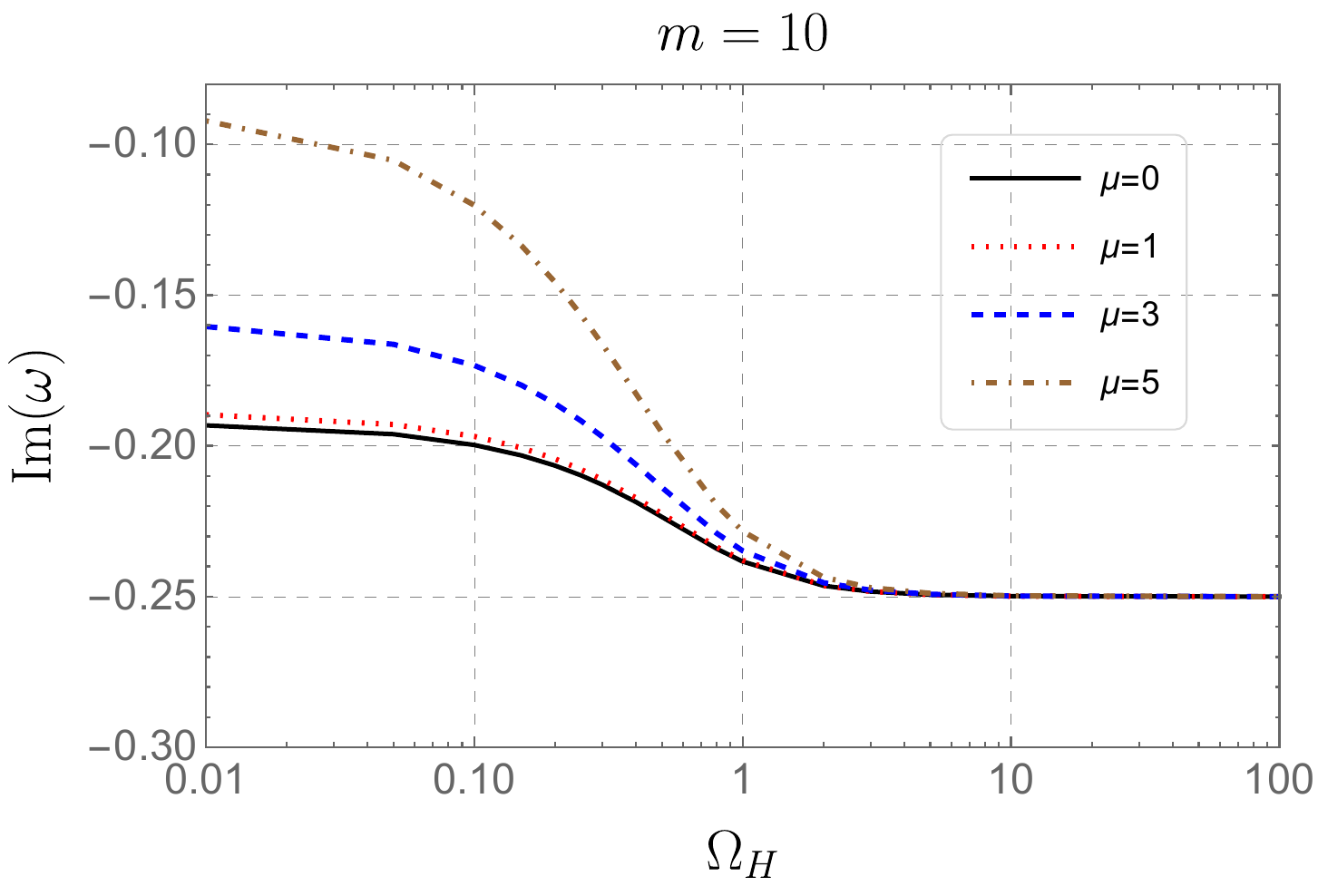}
\caption{The dependence of real part $\omega_R$ (left plots) and imaginary part $\omega_I$ (right plots) of fundamental QNF on $\Omega_H$ at $m=5$ and $m=10$ for various mass values.\label{fig2}}
\end{figure}

The dependence of real part and imaginary part of QNF given in Table~\ref{tab1} on the angular velocity  are separately plotted in Fig.~\ref{fig1} in order to have a more perspicuous demonstration of the features of QNF we have discussed above. In the following Table~\ref{tab2} and \ref{tab3} we respectively illustrate the QNF for $m=5$ and $m=10$, and the corresponding figures are shown in Fig.~\ref{fig2} from which we find that the QNF with higher positive winding number $m$ behaves qualitatively the same as the $m=1$ case, especially all the $\omega_I$ for the $m$ ranges from $m=1$ to $m=10$ approaches $\omega_I\approx-0.25$ when $\Omega_H$ is increasing to large value.

Based on the $\omega_R$ in Fig.~\ref{fig1} and \ref{fig2} are shown  by log-log plots, the slope corresponding to any given mass value $\mu$ at high $\Omega_H$ can be determined by linear regression
\begin{equation}
\frac{\Delta \ln \omega_R}{\Delta\ln \Omega_H}\approx1,
\end{equation}
which is dependent neither on  winding number nor perturbation field mass, and directly leads to
\begin{equation}
\ln\omega_R\approx\ln\Omega_H+\ln C_m. 
\end{equation}
By taking $\Omega_H=1$, the $m$ dependent constant $C_m$ is determined to be $C_m\approx m$, such that when $\Omega_H\to+\infty$ we have
\begin{equation}
 \omega_R\approx m\Omega_H, \quad \omega_I\approx-0.25, \quad m>0.
\end{equation}

%%%%%%%%%%%%%%%%%%%%%%%%%%%%%%%
\begin{table}[!htbp]
\centering
\caption*{$m=10$} 
\resizebox{\textwidth}{!}
{
        \begin{tabular}{ccccccc}
    \hline\hline
    $\mu$ &Method&  $\Omega_H=0$  &$\Omega_H=0.5$          & $\Omega_H=1$          & $\Omega_H=5$             & $\Omega_H=10$       \\
    \hline
    $0$ &WKB&     $ 3.84704 - 0.192464i $           & $ 6.92012 - 0.223568i $ & $ 11.148 - 0.23837i $ & $ 50.2488 - 0.249383i $    & $ 100.125 - 0.249844i $\\
        \cline{2-7}
        &CFM&     $ 3.84704 -0.192464i $            & $ 6.92012 -0.223568i $ & $ 11.148 -0.23837i $ & $ 50.2488 -0.249383i $    & $ 100.125 -0.249844i $\\
        \hline
    $1$ &WKB&     $ 3.89023 - 0.188852i $           & $ 6.94044 - 0.2225i $ & $ 11.1598 - 0.237976i $ & $ 50.2513 - 0.249365i $    & $ 100.126 - 0.249109i $\\
        \cline{2-7}
        &CFM&     $ 3.89023 -0.188852i $           & $ 6.94044 -0.2225i $ & $ 11.1598 -0.237976i $ & $ 50.2513 -0.249364i $    & $ 100.126 -0.249839i $\\
        \hline
    $3$ &WKB&     $ 4.24213 - 0.158906i $           & $ 7.10399 - 0.213812i $ & $ 11.2541 - 0.234804i $ & $ 50.2710 - 0.249249i $    & $ 100.137 - 0.252134i $\\
        \cline{2-7}
        &CFM&     $ 4.24213 -0.158906i $           & $ 7.10399 -0.213812i $ & $ 11.2541 -0.234804i $ & $ 50.2712 -0.249215i $    & $ 100.136 -0.249802i $\\
        \hline
    $5$ &WKB&     $ 4.99183 - 0.0887351i $           & $ 7.43701 - 0.195566i $ & $ 11.444 - 0.228342i $ & $ 50.3111 - 0.248916i $    & $ 100.156 - 0.249700i $\\
        \cline{2-7}
        &CFM&     $ 4.99183 -0.0887355i $           & $ 7.43701 -0.195566i $ & $ 11.444 -0.228342i $ & $ 50.3111 -0.248917i $    & $ 100.156 -0.249727i $\\
        \hline\hline
\end{tabular}
}
\caption{The fundamental QNF obtained by WKB and CFM  at $m=10$ for different mass values $\mu$ and angular velocity $\Omega_H$.}\label{tab3}
\end{table}
%%%%%%%%%%%%%%%%%%%%%%%%%%%%%%%

%%%%%%%%%%%%%%%%%%%%%%%%%%%%%%%
%%%%%%%%%%%%%%%%%%%%%%%%%%%%%%%

Now we focus our attention on the QNF for negative winding number $m<0$. In the negative $m$ case, our numerical methods are not capable of dealing with large $\Omega_H$, which is in contrast to $m>0$ case where the numerical methods work well for large $\Omega_H$.  On the other hand, the capability of WKB method to calculate precision massive QNF in this case is depressed. As a consequence,  we can only  rely on CFM to calculate QNF for $\Omega_H$ which is not large. 

\begin{figure}[thbp]
\centering
\includegraphics[height=2.4in,width=3.2in]{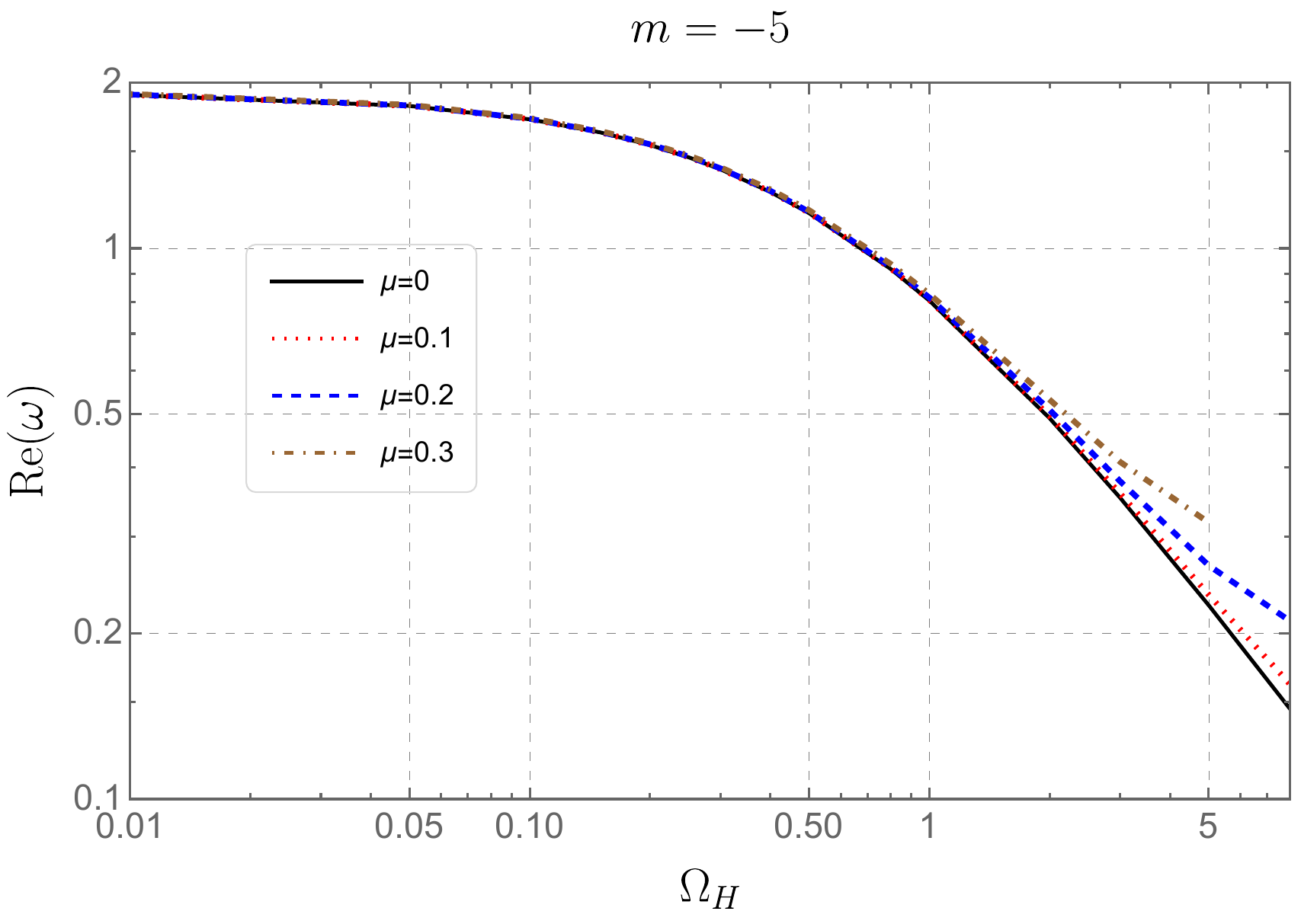}
\includegraphics[height=2.4in,width=3.2in]{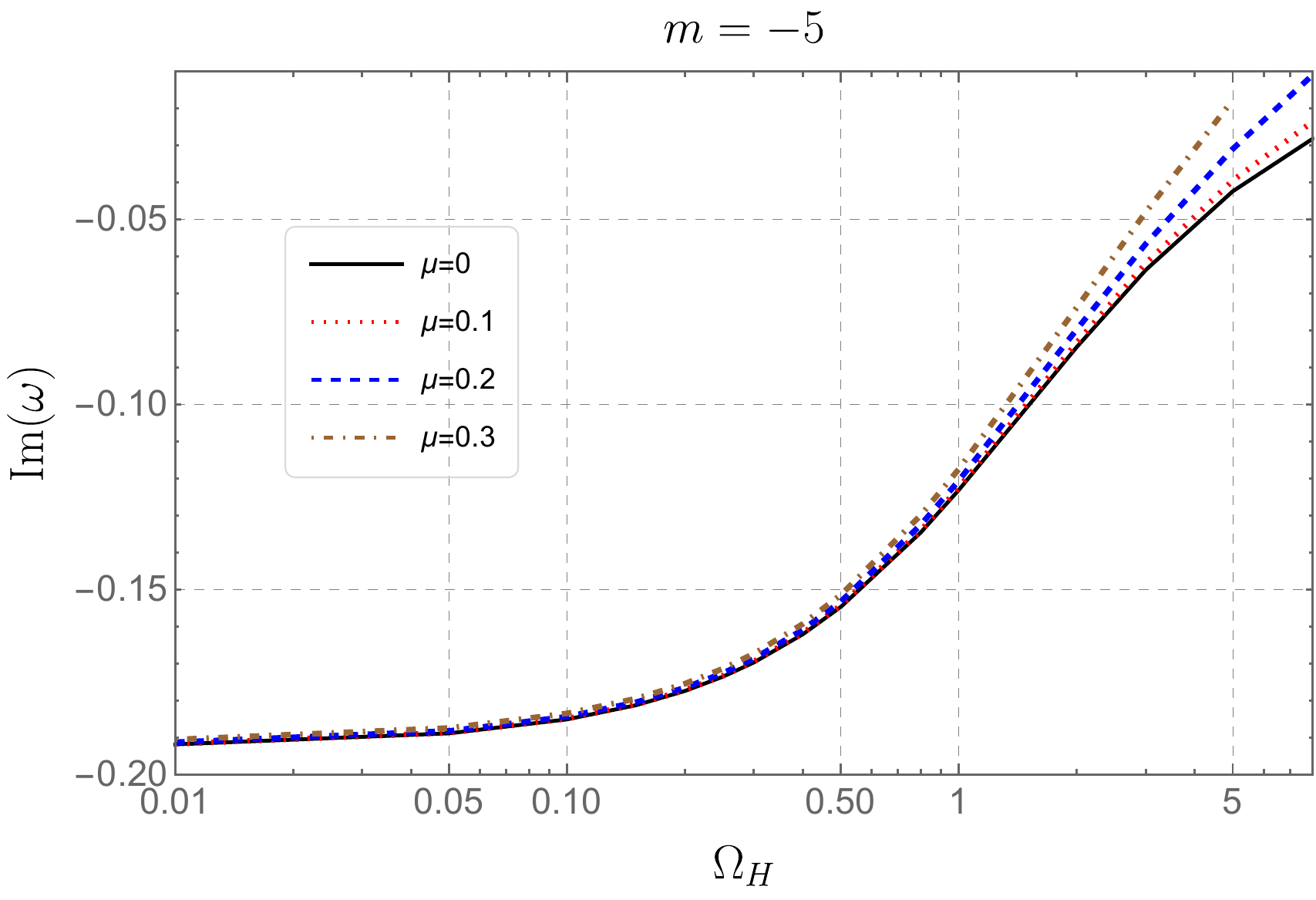}

\includegraphics[height=2.4in,width=3.2in]{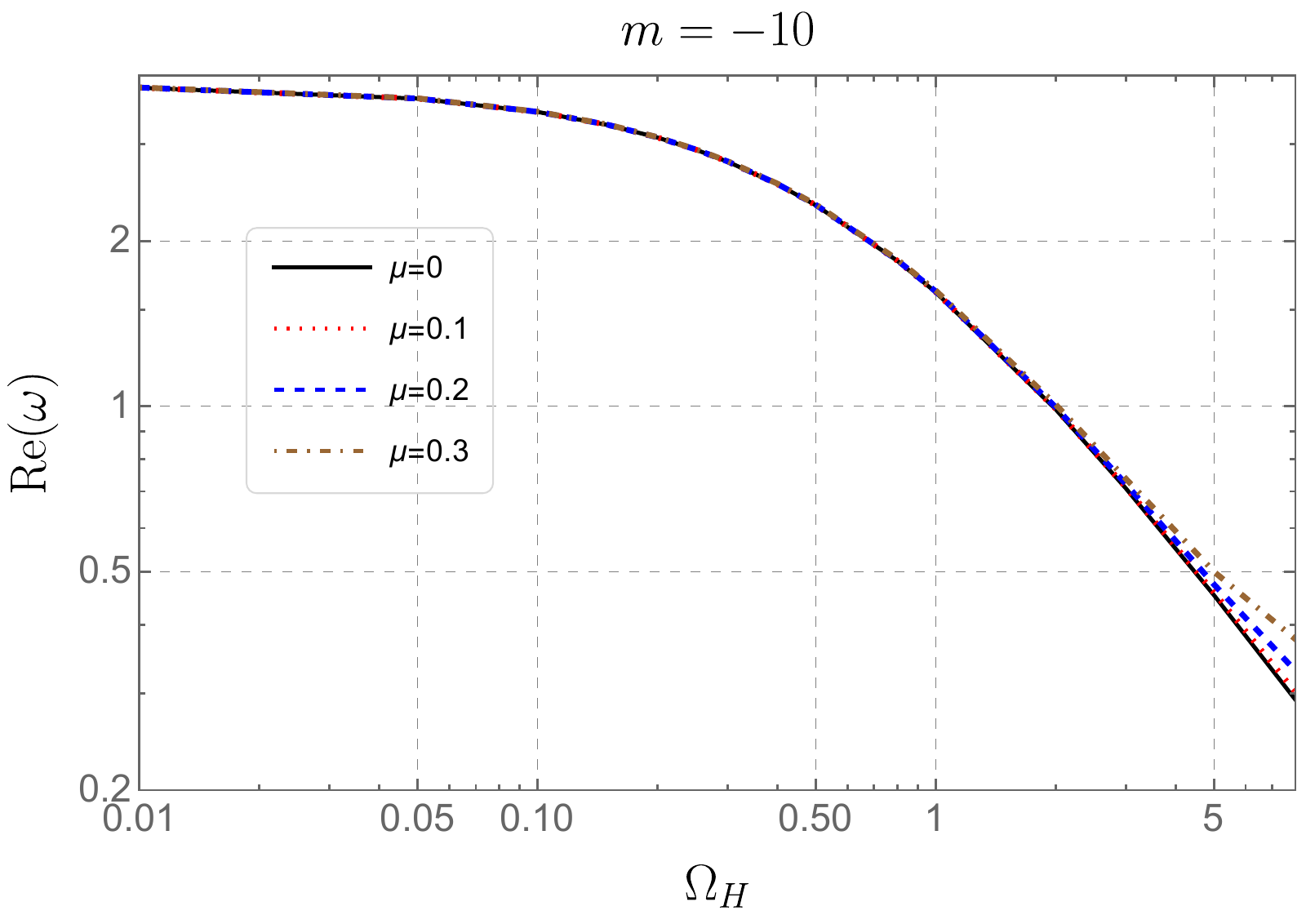}
\includegraphics[height=2.4in,width=3.2in]{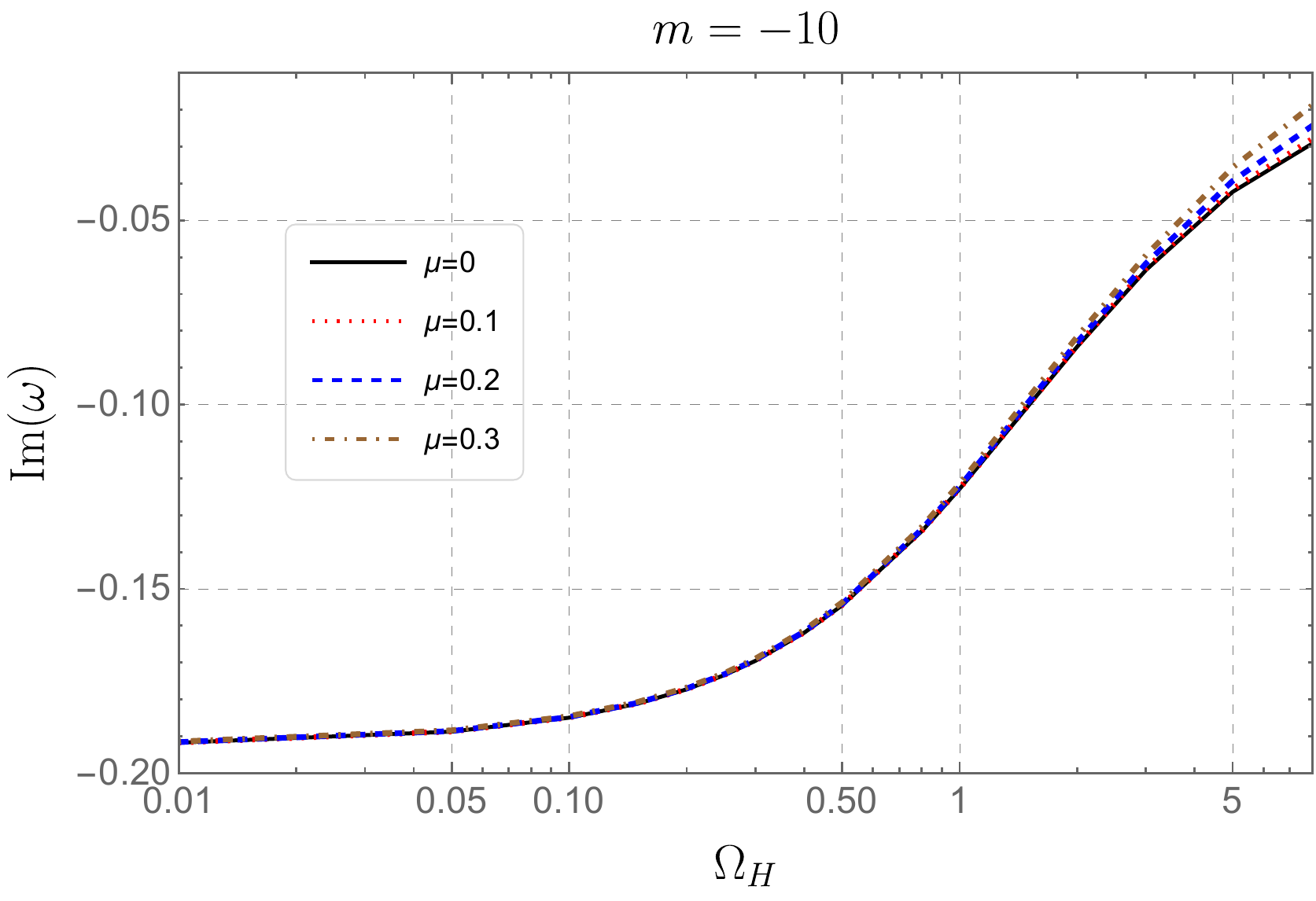}
\caption{The dependence of real part $\omega_R$ (left plots) and imaginary part $\omega_I$ (right plots) of fundamental QNF on $\Omega_H$ at $m=-5$ and $m=-10$ for various mass values.\label{fig3}}
\end{figure}

In Table~\ref{tab4}, we present QNF data for $m=-5$ under different mass and angular velocity values. One can observe that the QNF in current case shows a remarkably contrasting  features compared to positive $m$ case, as both $\omega_R$ and the magnitude of $\omega_I$ decreases with the grow of $\Omega_H$. However, the effects of perturbation field mass values $\mu$ on QNF are remained qualitatively identical to $m>0$ case, as when increasing $\mu$ the $\omega_R$ will become bigger while $\omega_I$ will be smaller. Hence we may conclude that the characteristics of the effects of $\Omega_H$ on QNF depend on the sign of winding number which will not change the way of how mass $\mu$ affects QNF. The data of QNF for $m=-10$ are listed in  Table~\ref{tab5} from which a qualitatively identical behaviors of QNF can be found.  This result may suggest that the fundamental QNF under all the negative $m$ has the same properties.  The corresponding plots of QNF in current two cases are demonstrated in Fig.~\ref{fig3} which gives clear illustrations of features of QNF for negative $m$. With the help of the plots, we find that the differences between QNF induced by different mass values will increase when $\Omega_H$ grows, this is another behavior differs from the $m>0$ case.

%%%%%%%%%%%%%%%%%%%%%%%%%%%%%%%
\begin{table}[!htbp]
\centering
\caption*{$m=-5$} 
\resizebox{\textwidth}{!}
{
        \begin{tabular}{ccccccc}
    \hline\hline
    $\mu$ &Method&  $\Omega_H=0$  &$\Omega_H=0.5$          & $\Omega_H=1$          & $\Omega_H=5$             & $\Omega_H=8$       \\
    \hline
    $0$   &CFM&     $ 1.92059 -0.192507i $   & $ 1.15837-0.154622i $ & $ 0.803536-0.123037i $ & $ 0.224911-0.0423650i $    & $ 0.145626 -0.0281561i $\\
        \hline
    $0.1$ &CFM&     $ 1.92144 -0.192364i $   & $ 1.15999 -0.154262i $ & $ 0.806051 -0.122409i $ & $ 0.234984 -0.0394395i $    & $ 0.161472 -0.0235405i $\\
        \hline
    $0.2$ &CFM&     $ 1.92399 -0.191935i $   & $ 1.16487 -0.153182i $ & $ 0.813602 -0.120527i $ & $ 0.265557 -0.0309244i $    & $ 0.210410 -0.0109232i $\\
        \hline
    $0.3$ &CFM&     $ 1.92825 -0.191219i $  & $ 1.17300 -0.151381i $ & $ 0.826214 -0.117392i $ & $ 0.317841 -0.0177205i $    & \textit{ non-convergence }\\
        \hline\hline
\end{tabular}
}
\caption{The fundamental QNF obtained by CFM  at $m=-5$ for different mass values $\mu$ and angular velocity $\Omega_H$. The\textit{ non-convergence } in the table means that our numerical method failed to give convergent results.}\label{tab4}
\end{table}
%%%%%%%%%%%%%%%%%%%%%%%%%%%%%%%

%%%%%%%%%%%%%%%%%%%%%%%%%%%%%%%
\begin{table}[!htbp]
\centering
\caption*{$m=-10$} 
\resizebox{\textwidth}{!}
{
        \begin{tabular}{ccccccc}
    \hline\hline
    $\mu$ &Method&  $\Omega_H=0$  &$\Omega_H=0.5$          & $\Omega_H=1$          & $\Omega_H=5$             & $\Omega_H=10$       \\
    \hline
    $0$   &CFM&     $ 3.84704 -0.192464i $   & $ 2.32298 -0.154473i $ & $ 1.61296 -0.122871i $ & $ 0.452634 -0.0422895i $    & $ 0.237419 -0.0229592i $\\
        \hline
    $0.1$ &CFM&     $ 3.84747 -0.192428i $   & $ 2.32381 -0.154383i $ & $ 1.61425 -0.122712i $ & $ 0.457837 -0.0415430i $    & $ 0.247562 -0.0214801i $\\
        \hline
    $0.2$ &CFM&     $ 3.84877 -0.192320i $   & $ 2.32630 -0.154110i $ & $ 1.61813 -0.122235i $ & $ 0.473465 -0.0393243i $    & $ 0.278097 -0.0172152i $\\
        \hline
    $0.3$ &CFM&     $ 3.85092 -0.192140i $  & $ 2.33045 -0.153655i $ & $ 1.62459 -0.121441i $ & $ 0.499575 -0.0356964i $    & $ 0.329391 -0.0107323i $\\
        \hline\hline
\end{tabular}
}
\caption{The fundamental QNF obtained by CFM  at $m=-10$ for different mass values $\mu$ and angular velocity $\Omega_H$. }\label{tab5}
\end{table}
%%%%%%%%%%%%%%%%%%%%%%%%%%%%%%%

%%%%%%%%%%%%%%%%%%%%%%%%%%%%%%%
%%%%%%%%%%%%%%%%%%%%%%%%%%%%%%%

\begin{figure}[thbp]
\centering
\includegraphics[height=2.4in,width=3.2in]{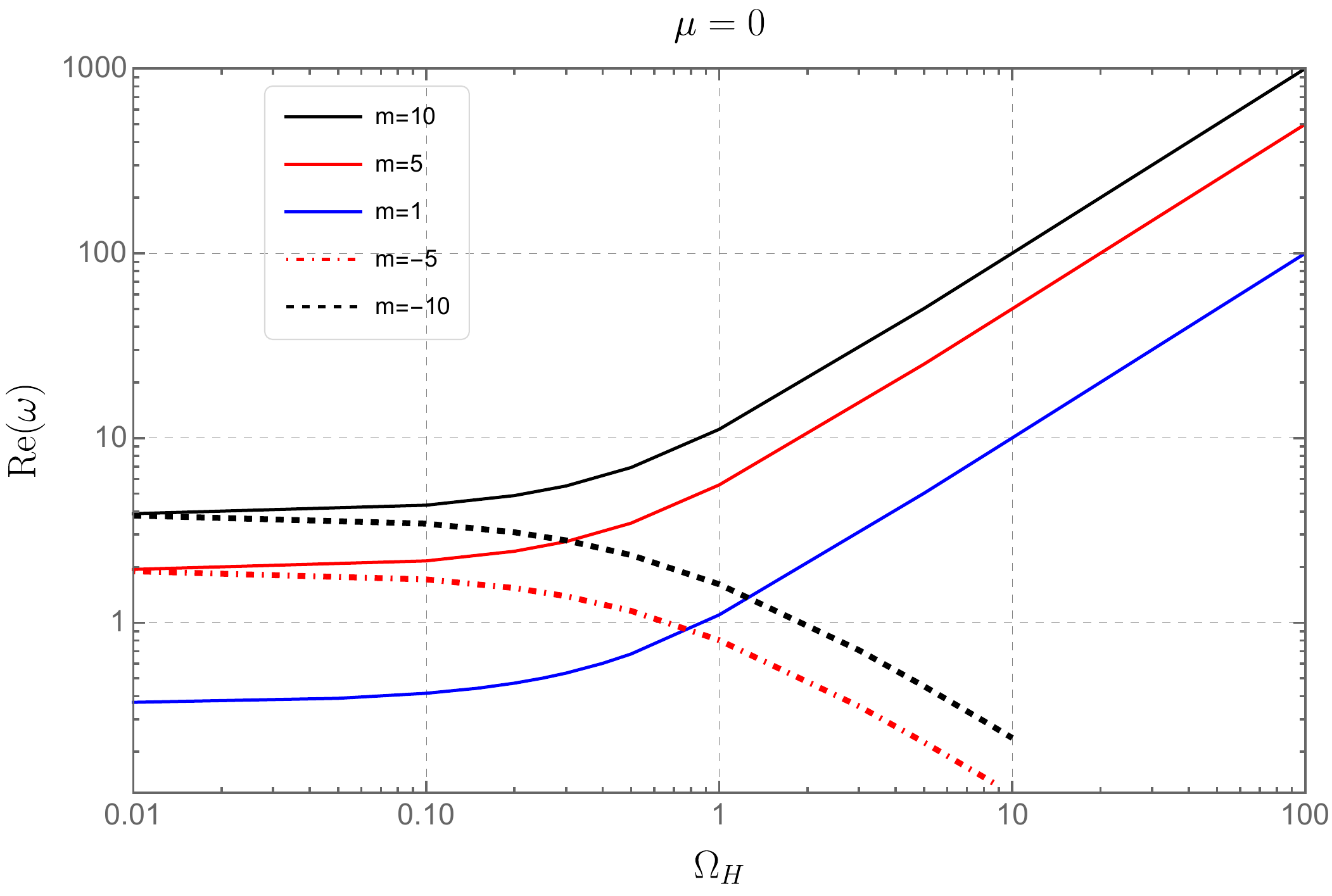}
\includegraphics[height=2.4in,width=3.2in]{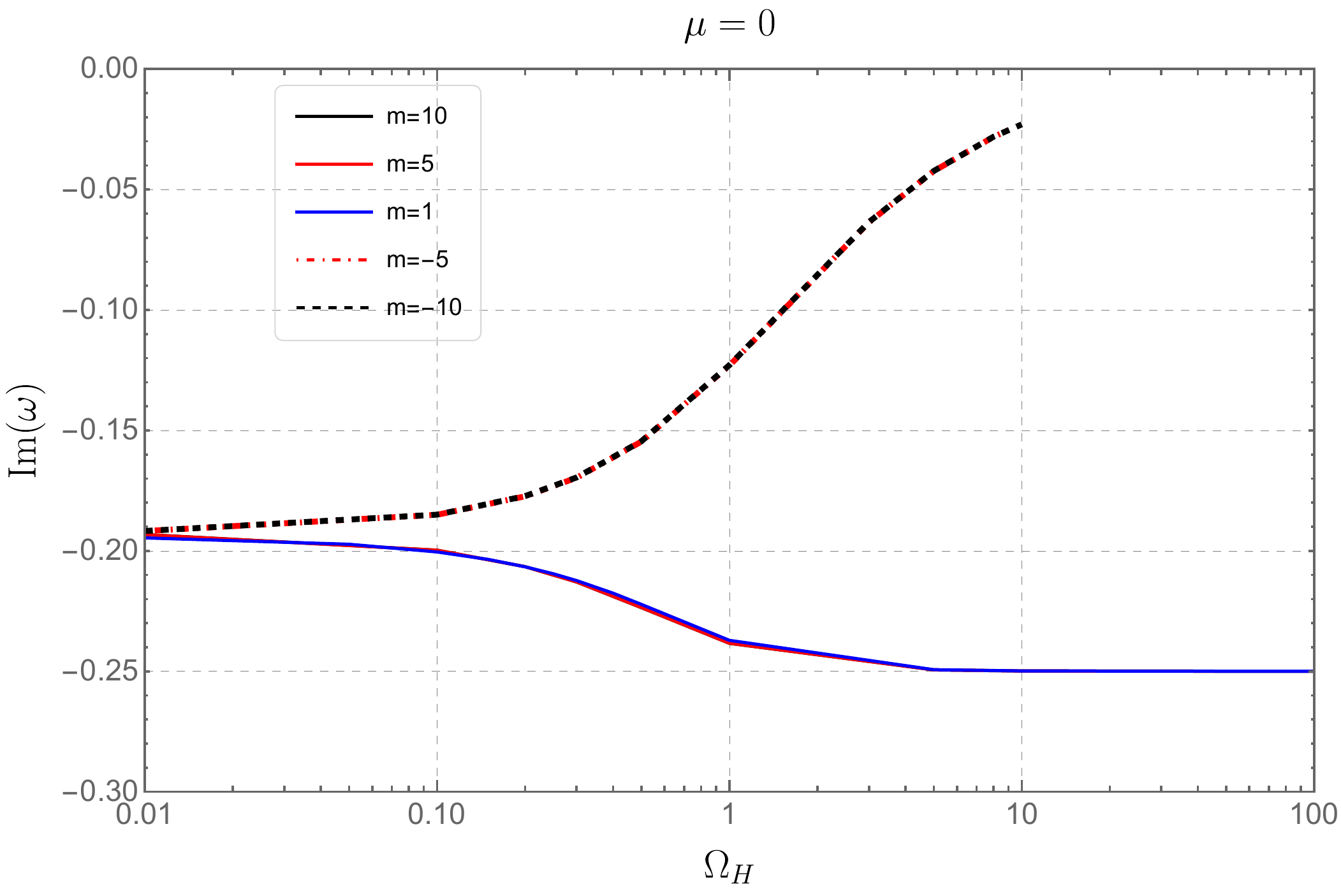}

\includegraphics[height=2.4in,width=3.2in]{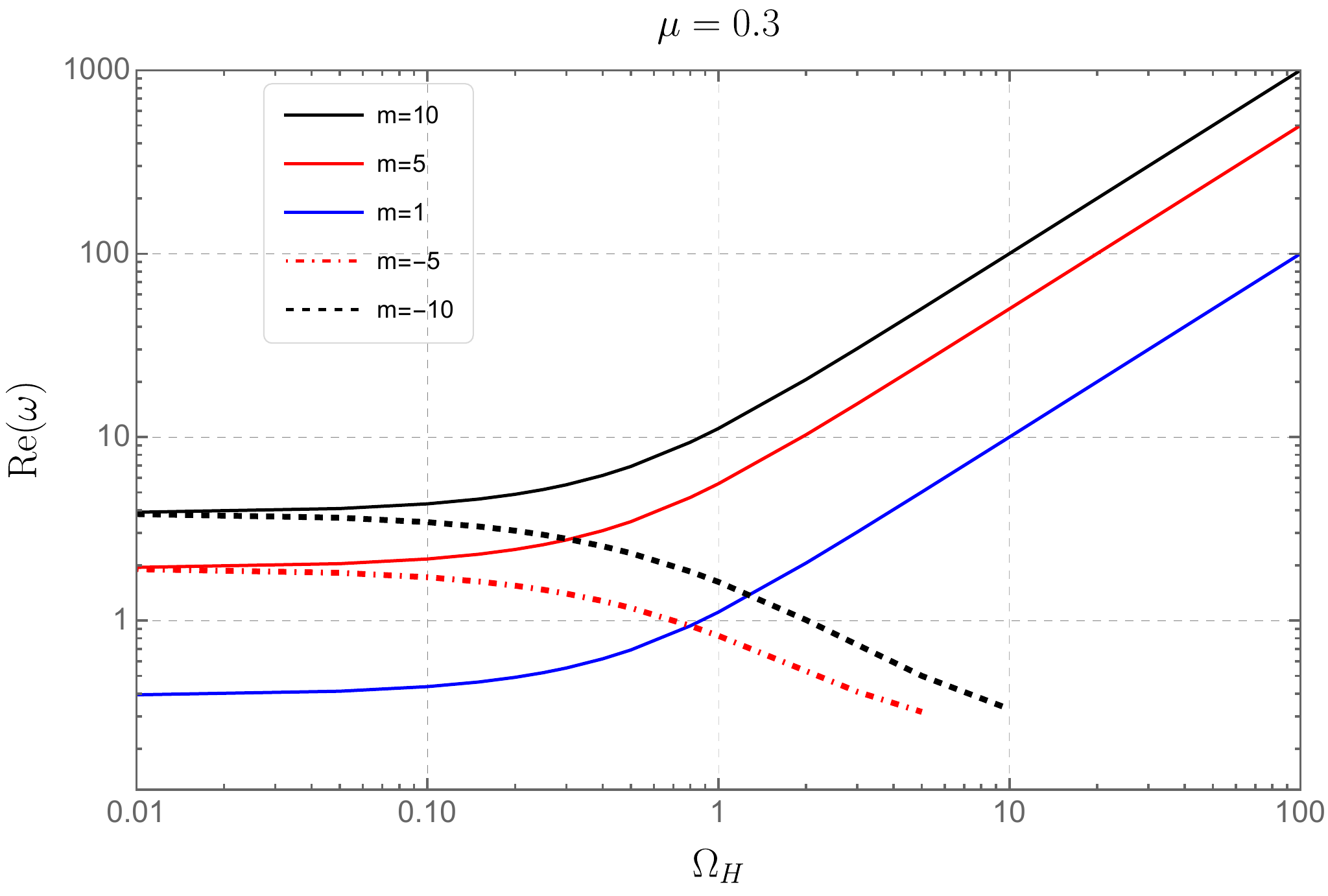}
\includegraphics[height=2.4in,width=3.2in]{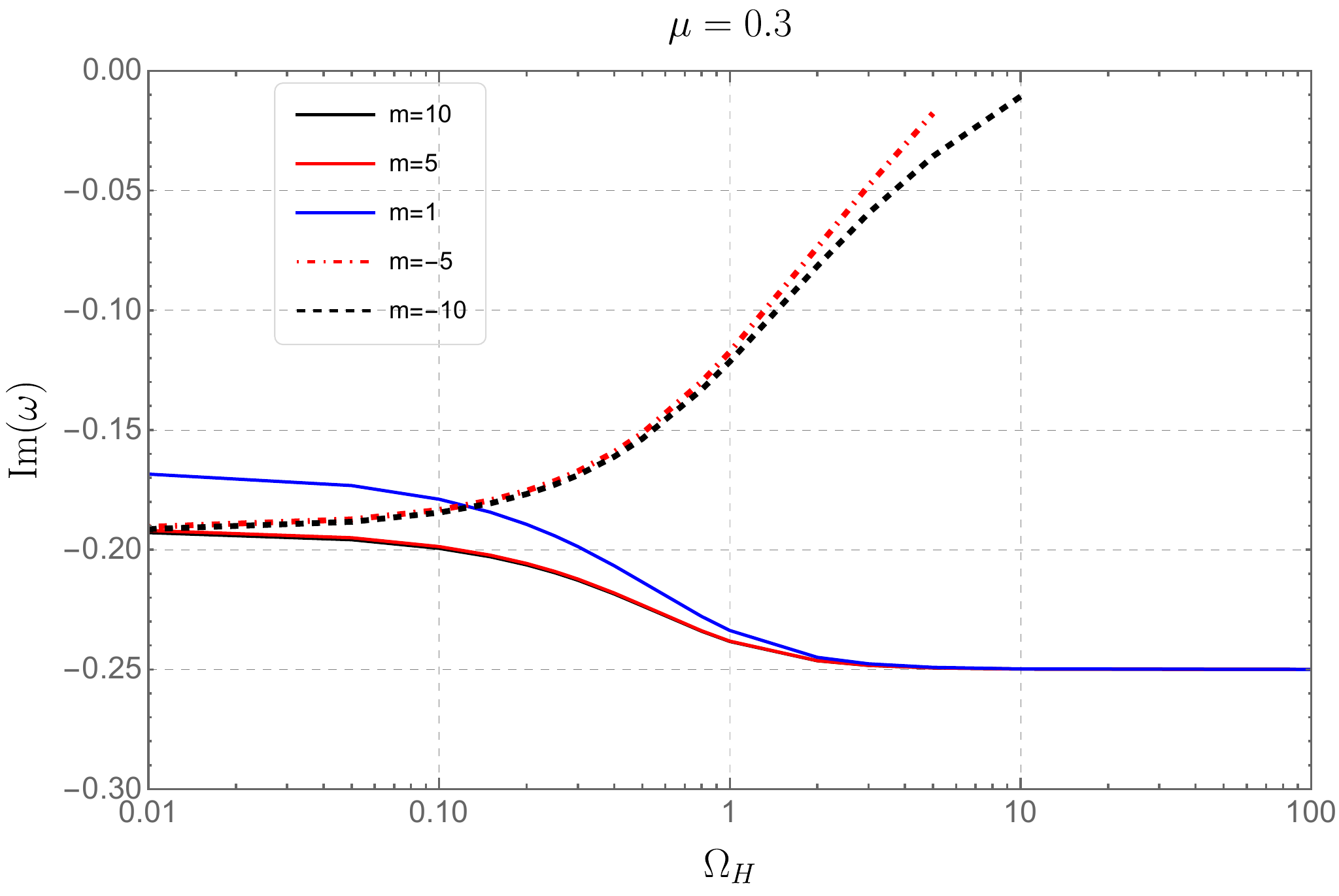}
\caption{The comparison of fundamental QNF between different winding numbers $m$ for scalar field mass $\mu=0$ (plots on the top) and $\mu=0.3$ (plots on the bottom).\label{fig4}}
\end{figure}

As the last part of this section, we would like to compare the QNF among different wingding numbers in order to disclose how the QNF is affected by $m$. The comparisons are made in Fig.~\ref{fig4} where we plot the QNF among $m=-10$ to $m=10$ by separately fixing $\mu=0$ and $\mu=0.3$ in each plot. The plots of $\omega_R$ and $\omega_I$ are placed on the left side and right side of Fig.~\ref{fig4}, respectively. For the $\omega_R$, it can be found that different mass values $\mu$ do not have noticeable influences on the behaviors of $\omega_R$.  Regarding the influences of the winding number on $\omega_R$, it is clearly to see that among the positive $m$, the QNF with a larger winding number always has a higher $\omega_R$. When it comes to negative $m$,  the larger $\omega_R$ comes from the modes with higher magnitude $|m|$. The plots for $\omega_I$ under the two different mass values manifest obvious unique features. In the $\mu=0$ plot, the $\omega_I$  for all $m>0$ coincides and forms a single branch, while another branch is formed by $\omega_I$ for all wingding numbers $m<0$. This result means that the $\omega_I$ of fundamental QNF with mass $\mu=0$ depends on the sign of winding number whose magnitude can just mildly affect it. However, for the $\omega_I$ of massive perturbation field, as shown in our plot in which  $\mu=0.3$, this kind of degeneracy is broken. When we change $\mu=0$ into $\mu=0.3$, the original two branches start to diverge. This divergency in $m<0$ branch happens and becomes more pronounced   when increasing $\Omega_H$ to larger values. In contrast, in the $m>0$ branch, the divergence occurs at small $\Omega_H$ and it starts to  disappear if we keep increasing  $\Omega_H$ to large values.

%%%%%%%%%%%%%%%%%%%%%%%%%%%%%%%
%%%%%%%%%%%%%%%%%%%%%%%%%%%%%%%

\section{Quasi-Resonance}\label{sec5}

In this section we discuss an interesting phenomenon uncovered in the spectrum of  massive QNMs. In the calculation of QNF of massive perturbation field, people have  found that the $\omega_I$ of QNF will increase and gradually approaches zero when we improve  the  mass $\mu$ to some value, while the $\omega_R$ remains to be nonvanishing. As a consequence, the QNMs can be arbitrarily long-lived due to the $\omega_I$ which represents damping rate of QNMs can be sufficiently small. This phenomenon is called quasi-resonance which was first observed in \cite{Ohashi:2004wr} where the authors studied the massive scalar field  perturbation in the Reissner-Nordström black hole spacetime. Subsequently, it was realized in \cite{Konoplya:2004wg} that  quasi-resonance can exist only if the effective potential is nonzero at infinity. Note that the effective potential in the master equation of the massive scalar perturbation in present work, therefore we would like to learn whether the quasi-resonance can exist in our analog rotating black hole system. To this end, we need to observe how the QNF behaves when we  change the value of perturbation field mass.

%%%%%%%%%%%%%%%%%%%%%%%%%%%%%%%
\begin{figure}[thbp]
\centering
\includegraphics[height=2.4in,width=3.2in]{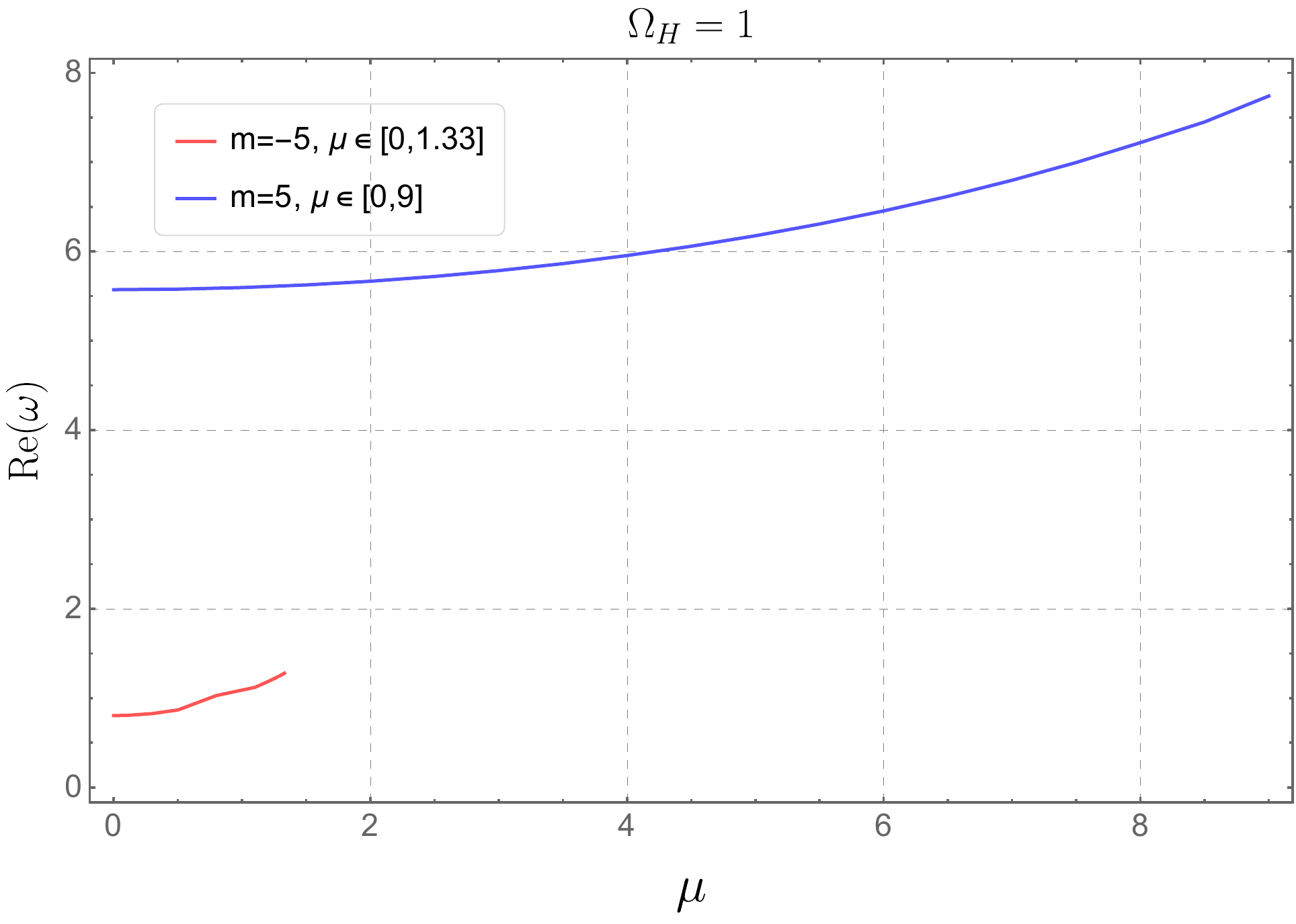}
\includegraphics[height=2.4in,width=3.2in]{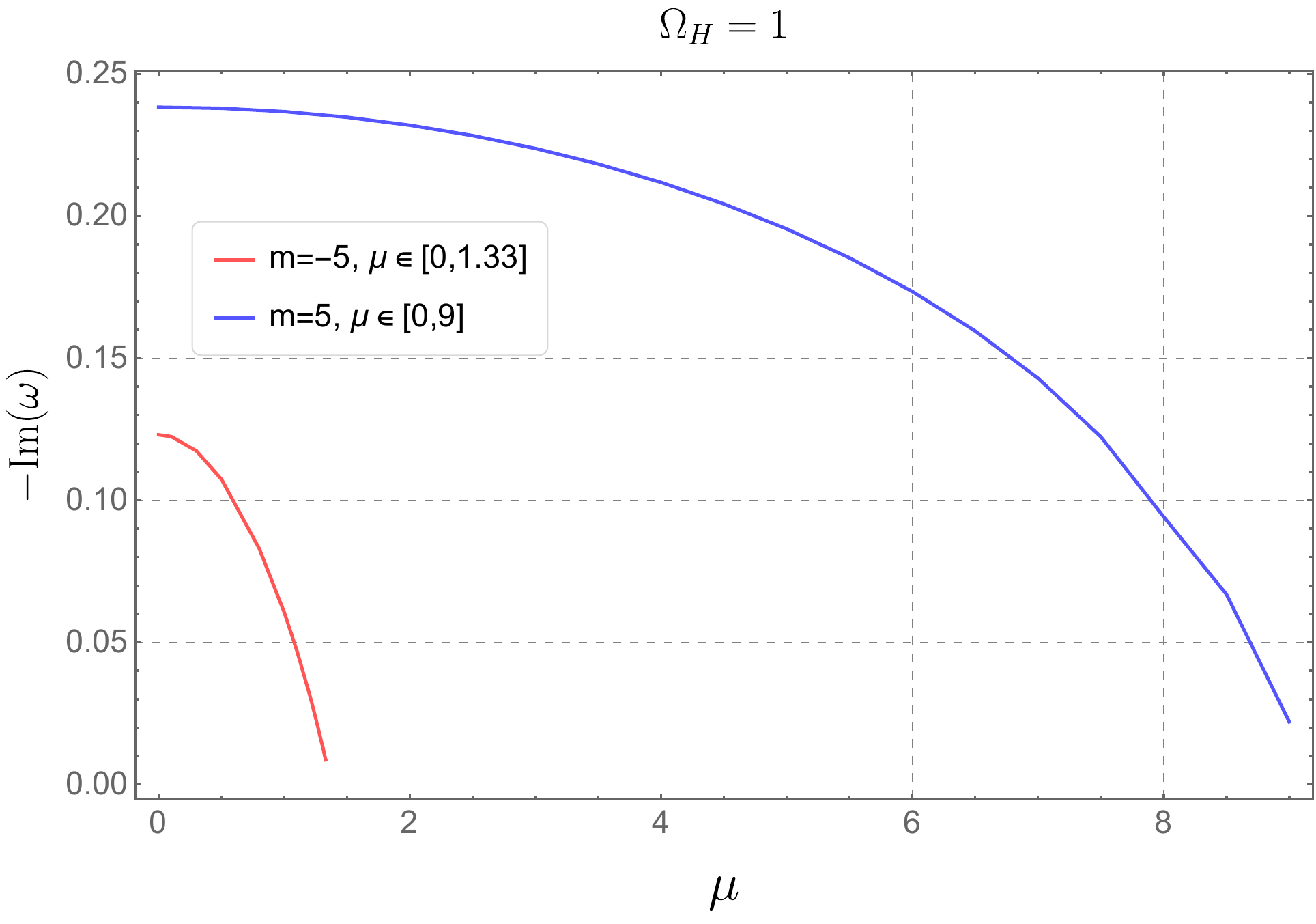}
\caption{The dependence  of fundamental QNF at $\Omega_H=1$ for different winding numbers $m$ on scalar field mass $\mu$.\label{fig5}}
\end{figure}
%%%%%%%%%%%%%%%%%%%%%%%%%%%%%%%

\begin{figure}[thbp]
\centering
\includegraphics[height=2.5in,width=3.4in]{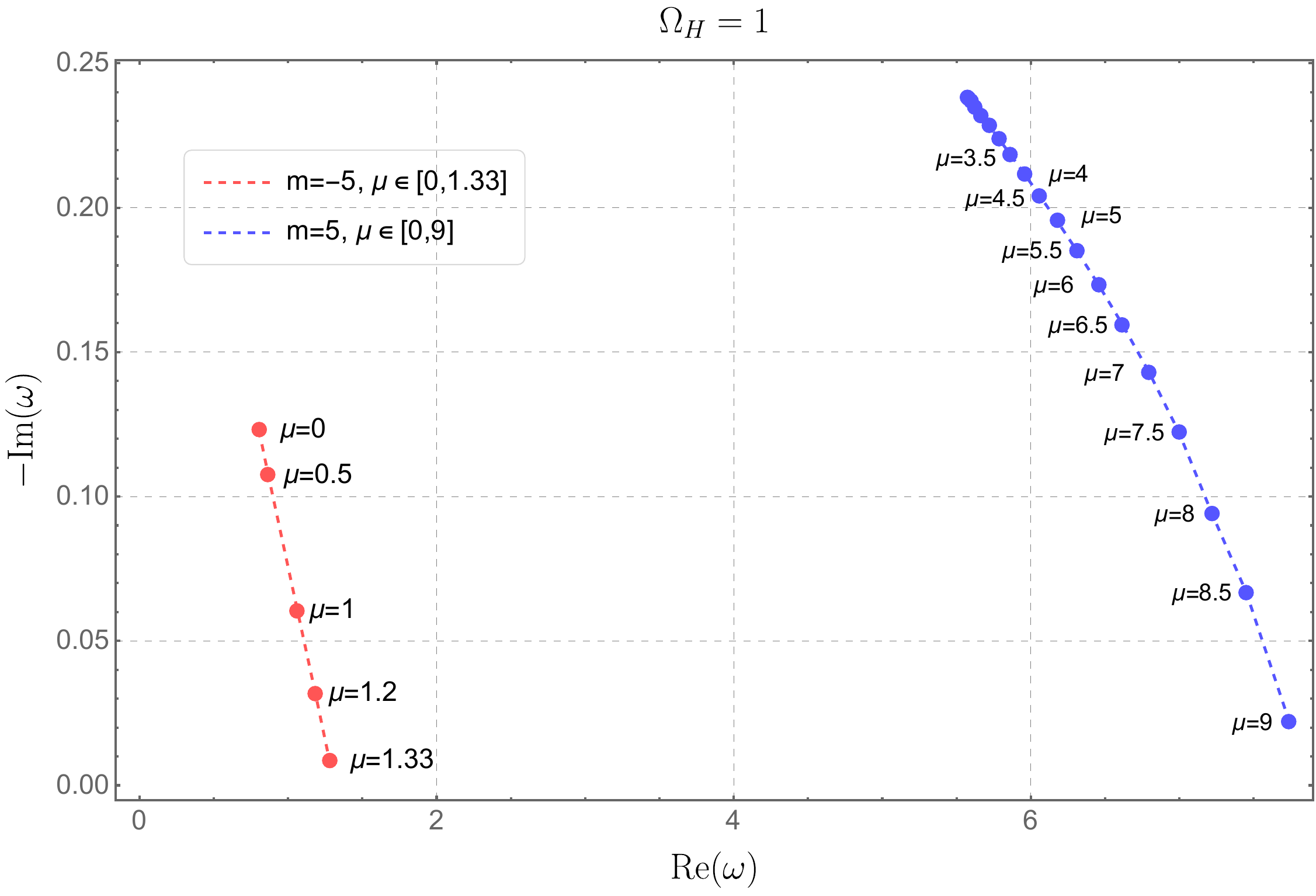}
\caption{The fundamental QNF demonstrated in complex plane  for different mass values $\mu$ and winding number $m$ at $\Omega_H=1$.\label{fig6}}
\end{figure}

In Fig.~\ref{fig5} we demonstrate the dependence of real and imaginary part  of fundamental QNF for $m=\pm 5$ at $\Omega_H=1$ on the perturbation field mass $\mu$. We should point out that we calculated QNF for $m=-5$ in a small range of $\mu$ compared to $m=5$, since our numerical method can  yield  convergent results only for comparatively small $\mu$ in the context of negative winding number. For both positive and negative $m$, this figure shows that when increasing $\mu$, the $\omega_R$ will grow while the magnitude of $\omega_I$ decreases meanwhile with a trend to get close to zero. The zero-approaching tendency of $\omega_I$ suggests that a sufficient small damping rate can be achieved  if we  further increase $\mu$ to some certain value, but the numerical method start to work poorly which stops us here.  In this sense, we may claim that the quasi-resonance can exist in present analog rotating black hole system. Another property reflected by the plots is that the QNF with negative wingding number seem to be more sensitive to the change of $\mu$, as the slope of the curves for $m=-5$ seems to be higher than that of $m=5$. On the other hand, we combine the two plots together in Fig.~\ref{fig5} to get the behaviors of QNF on the complex plane, as shown in Fig.~\ref{fig6} which gives a more clear and straightforward illustration of how QNF behaves under the variation of mass $\mu$.

%%%%%%%%%%%%%%%%%%%%%%%%%%%%%%%
%%%%%%%%%%%%%%%%%%%%%%%%%%%%%%%

\section{Conclusions and Discussions}\label{sec6}

In this paper, we have investigated the fundamental QNF of massive scalar field perturbation in the analog rotating black hole spacetime which is established by photon-fluid system. The  QNF with opposite sign of winding number is found to behave rather differently, while the modes with the same sign of winding number have qualitatively similar properties. If winding number $m>0$, both the $\omega_R$ and the magnitude of $\omega_I$ which is negative  increase with the $\omega_H$. When $\omega_H$ is large enough, the effects of the mass values of perturbation field on the QNF will be suppressed and all the QNF behaves identically, as the $\omega_R$ will asymptotically behave as $\omega_R\approx m\Omega_H$, and the $\omega_I$ will approach a constant $\omega_I\approx-0.25$. In winding number $m<0$ case, the increase  of $\Omega_H$ will lead to the  decrease of  both $\omega_R$ and magnitude of $\omega_I$, which is  contrary  to $m>0$ case. The influences of winding number on QNF are also analyzed. We find that among the positive winding numbers, the QNMs with bigger $m$ always has a larger $\omega_R$, and this result is reversed among the QNMs with $m<0$. We also note that the perturbation field mass seems not to change the behaviors of $\omega_R$ among different winding numbers, while it has been found to have noticeable influences on $\omega_I$. In the $\mu=0$ case, we find a kind of degeneracy, which manifests as all the $\omega_I$ with the same sign of winding number almost coincide with each other. As a consequence, we can only find two branches of curves standing for positive and negative winding numbers, respectively. This phenomenon indicates that for the massless QNF,  only the sign of winding number can influence  $\omega_I$ which is almost independent of the magnitude of $m$. However, when the perturbation field mass $\mu$ becomes nonzero, we find that the degeneracy is broken, as the original single branch will diverge. Note that the divergence will be enhanced when  increasing $\Omega_H$ for $m<0$ QNMs, and for $m>0$ it will be suppressed and eventually converge. Despite the QNMs with opposite sign of winding number usually  have very contrasting characteristics,  the perturbation field mass $\mu$ can influence the QNF with all different $m$ in the same way: it increases the $\omega_R$ and reduces the magnitude of $\omega_I$.

Besides the conventional properties of QNMs, we also studied the quasi-resonance which is an interesting feature of the massive QNMs. The quasi-resonance is a kind of QNF whose real part $\omega_R$ remains nonvanishing while the imaginary part $\omega_I$ can approach zero which suggests the existence of arbitrarily long-lived QNMs. We find that the QNF in complex plane will quickly move toward the real axis as we have shown in Fig.~\ref{fig6}, which may indicate the existence of  quasi-resonance in current analog rotating black hole system. It is worth pointing out that the quasi-resonance is favored in the study of black hole physics through analog gravity, since its long damping time will make it less hard for us to successfully detect QNMs in experiment. As an aside, note that this analog rotating  black hole has been experimentally constructed in~\cite{Vocke2018}, and the recent work in \cite{Solidoro:2024yxi} has demonstrated  that the QNMs can indeed be easily excited  in finite-sized  experimental settings of analog gravity. In light of these two developments, we hopefully expect  that the successful detection of the QNMs in  analog gravity model by photon-fluid system can be realized, and thus  bringing us a more promising future of studying the physics in the black holes spacetime by analogy.

\begin{acknowledgments}
H.L. is grateful to  Yanfei for her unwavering  support to his career. This work is supported by the National Natural Science Foundation of China under Grant No.12305071. We also gratefully acknowledge the financial support from Brazilian agencies Funda\c{c}\~ao de Amparo \`a Pesquisa do Estado de S\~ao Paulo (FAPESP), Funda\c{c}\~ao de Amparo \`a Pesquisa do Estado do Rio de Janeiro (FAPERJ), Conselho Nacional de Desenvolvimento Cient\'{\i}fico e Tecnol\'ogico (CNPq), and Coordena\c{c}\~ao de Aperfei\c{c}oamento de Pessoal de N\'ivel Superior (CAPES).
\end{acknowledgments}

\appendix
\section{The derivation of the master equation}\label{app1}
To obtain the radial wave equation of the perturbation field, we perform a separation of variable to $\rho_1$, as
\begin{equation}
	\rho_1(t,r,\theta)=R(r)e^{-i(\omega t -m \theta)},\label{eq1}
\end{equation}
where integer $m$ is called the winding number. We substitute Eq.~\eqref{eq1} into Klein-Gordon equation to get
\begin{align}
	&\frac{d^2R(r)}{dr^2}+P(r)\frac{dR(r)}{dr}+Q(r)R(r)=0,\label{eq2}\\
	&P_1(r)=\frac{1}{r-r_H},\\
	&Q_1(r)=\frac{r^3\left(r_H \mu^2+r\left(-\mu^2+\omega^2\right)\right)-2 m r^2 r_H^2 \omega \Omega_H+m^2\left(-r^2+r r_H+r_H^4 \Omega_H^2\right)}{r^2(r-r_H)^2}.
\end{align}
Now we set
\begin{equation}
R(r)=G(r)\Psi(r),
\end{equation}
and then introduce a new coordinate $r_\ast$ defined by
\begin{equation}
\frac{dr_\ast}{dr}=\Delta(r).
\end{equation}
When working in this new coordinate, Eq.~\eqref{eq2} will be transformed into 
\begin{equation}
	G(r)\Delta^2(r)\frac{d^2\Psi(r_\ast)}{dr_\ast^2}+P_2(r)\frac{d\Psi(r_\ast)}{dr_\ast}+Q_2(r)\Psi(r_\ast)=0,\label{eq3}
\end{equation}
where
\begin{align}
	P_2(r)=\frac{G(r)\Delta(r)}{r-r_H}+2\Delta(r)\frac{dG(r)}{dr}+G(r)\frac{d\Delta(r)}{dr},\\
	Q_2(r)=\frac{d^2G(r)}{dr^2}+P_1(r)\frac{dG(r)}{dr}+Q_1(r)G(r).
\end{align}
In order to obtain the Schr\"{o}dinger-like equation, the coefficient $P_2(r)$ of $d\Psi(r_\ast)/dr_\ast$ should vanish
\begin{equation}
	\frac{G(r)\Delta(r)}{r-r_H}+2\Delta(r)\frac{dG(r)}{dr}+G(r)\frac{d\Delta(r)}{dr}=0.
\end{equation}
This equation can be easily solved by following solution
\begin{equation}
G(r)=\frac{1}{\sqrt{(r-r_H)\Delta(r)}}.	
\end{equation}
In order to get Schrodinger-like master equation and map the radial coordinate $r$ to turtle coordinate $r_\ast$ which runs over $(-\infty,+\infty)$, we take the following definition of $\Delta(r)$
\begin{equation}
\Delta(r)=\left(1-\frac{r_H}{r}\right)^{-1}.	
\end{equation}
Substituting $G(r)$ and $\Delta(r)$ into Eq.~\eqref{eq3}, we finally arrive at the master equation Eq.~\eqref{mastereq} of the perturbation field.

\bibliographystyle{JHEP}
%\bibliographystyle{apsrev4-1}
%注意tex文件名不能有空格否则参考文献识别不出来！！！！！！！！！！！！！！！！
\bibliography{References_Analog_BH1,References_Analog_BH2}

\end{document}